\documentclass[a4paper, 11pt]{article}
\pdfoutput=1

\usepackage{jheppub}
\usepackage{afterpage}

\newcommand{\Dmq}{\Delta m^2}
\newcommand{\eVq}{\ensuremath{\text{eV}^2}}
\newcommand{\Nuc}[2]{\ensuremath{\mbox{}^{#1}\text{#2}}}

\renewcommand{\Im}{\mathop{\rm Im}}
\newenvironment{pagefigure}{\begin{figure}[!p]}{\afterpage\clearpage\end{figure}}

\allowdisplaybreaks

\title{Updated fit to three neutrino mixing: status of leptonic CP violation}

\author[a,b]{M.~C.~Gonzalez-Garcia,}
\affiliation[a]{Instituci\'o Catalana de Recerca i Estudis
  Avan\c{c}ats (ICREA), Departament d'Estructura i Constituents de la
  Mat\`eria and Institut de Ciencies del Cosmos, Universitat de
  Barcelona, Diagonal 647, E-08028 Barcelona, Spain}
\affiliation[b]{C.N.~Yang Institute for Theoretical Physics, State
  University of New York at Stony Brook, Stony Brook, NY 11794-3840,
  USA}
\emailAdd{maria.gonzalez-garcia@stonybrook.edu}

\author[c]{Michele Maltoni,}
\affiliation[c]{Instituto de F\'{\i}sica Te\'orica UAM/CSIC, Calle de
  Nicol\'as Cabrera 13--15, Universidad Aut\'onoma de Madrid,
  Cantoblanco, E-28049 Madrid, Spain}
\emailAdd{michele.maltoni@csic.es}

\author[d]{Thomas Schwetz}
\affiliation[d]{Oskar Klein Centre for Cosmoparticle Physics,
  Department of Physics, Stockholm University, SE-10691 Stockholm,
  Sweden}
\emailAdd{schwetz@fysik.su.se}

\abstract{We present a global analysis of solar, atmospheric, reactor
  and accelerator neutrino data in the framework of three-neutrino
  oscillations based on data available in summer 2014. We provide the
  allowed ranges of the six oscillation parameters and show that their
  determination is stable with respect to uncertainties related to
  reactor neutrino and solar neutrino flux predictions. We find that
  the maximal possible value of the Jarlskog invariant in the lepton
  sector is $0.0329 \pm 0.0009$ ($\mathrel{\pm} 0.0027$) at the
  $1\sigma$ ($3\sigma$) level and we use leptonic unitarity triangles
  to illustrate the ability of global oscillation data to obtain
  information on CP violation.  We discuss ``tendencies and tensions''
  of the global fit related to the octant of $\theta_{23}$ as well as
  the CP violating phase $\delta_\text{CP}$.  The favored values of
  $\delta_\text{CP}$ are around $3\pi/2$ while values around $\pi/2$
  are disfavored at about $\Delta\chi^2 \simeq 6$.  We comment on the
  non-trivial task to assign a confidence level to this $\Delta\chi^2$
  value by performing a Monte Carlo study of T2K data.}

\preprint{IFT-UAM/CSIC-14-095, YITP-SB-14-31}

\keywords{neutrino oscillations, solar and atmospheric neutrinos}

\begin{document}

\maketitle

\section{Introduction}
\label{sec:intro}

Thanks to remarkable discoveries by a number of neutrino oscillation
experiments it is now an established fact that neutrinos have mass and
leptonic flavors are not symmetries of Nature~\cite{Pontecorvo:1967fh,
  Gribov:1968kq}, see Ref.~\cite{GonzalezGarcia:2007ib} for an
overview.
Ignoring controversial indications for the existence of neutrino mass
states at the eV scale (see Ref.~\cite{Kopp:2013vaa} and references
therein) a consistent description of global data on neutrino
oscillations is possible by assuming mixing among the three known
neutrinos ($\nu_e$, $\nu_\mu$, $\nu_\tau$), which can be expressed as
quantum superpositions of three massive states $\nu_i$ ($i=1,2,3$)
with masses $m_i$.  This implies the presence of a leptonic mixing
matrix in the weak charged current interactions~\cite{Maki:1962mu,
  Kobayashi:1973fv} which can be parametrized as~\cite{PDG}:
\begin{equation}
  \label{eq:matrix}
  U =
  \begin{pmatrix}
    c_{12} c_{13}
    & s_{12} c_{13}
    & s_{13} e^{-i\delta_\text{CP}}
    \\
    - s_{12} c_{23} - c_{12} s_{13} s_{23} e^{i\delta_\text{CP}}
    & \hphantom{+} c_{12} c_{23} - s_{12} s_{13} s_{23}
    e^{i\delta_\text{CP}}
    & c_{13} s_{23}
    \\
    \hphantom{+} s_{12} s_{23} - c_{12} s_{13} c_{23} e^{i\delta_\text{CP}}
    & - c_{12} s_{23} - s_{12} s_{13} c_{23} e^{i\delta_\text{CP}}
    & c_{13} c_{23}
  \end{pmatrix},
\end{equation}
where $c_{ij} \equiv \cos\theta_{ij}$ and $s_{ij} \equiv
\sin\theta_{ij}$.  In addition to the Dirac-type phase
$\delta_\text{CP}$, analogous to that of the quark sector, there may
also be two physical phases associated to a possible Majorana
character of neutrinos, which however are not relevant for neutrino
oscillations~\cite{Bilenky:1980cx, Langacker:1986jv} and are therefore
omitted in the present work.  Given the observed hierarchy between the
solar and atmospheric mass-squared splittings there are two possible
non-equivalent orderings for the mass eigenvalues, which are
conventionally chosen as
\begin{align}
  \label{eq:normal}
  \Dmq_{21} &\ll \hphantom{+} (\Dmq_{32} \simeq \Dmq_{31} > 0) \,;
  \\
  \label{eq:inverted}
  \Dmq_{21} &\ll -(\Dmq_{31} \simeq \Dmq_{32} < 0) \,,
\end{align}
with $\Dmq_{ij} \equiv m_i^2 - m_j^2$.  As it is customary we refer to
the first option, Eq.~\eqref{eq:normal}, as Normal Ordering (NO), and
to the second one, Eq.~\eqref{eq:inverted}, as Inverted Ordering (IO);
in this form they correspond to the two possible choices of the sign
of $\Dmq_{31}$. In this convention the angles $\theta_{ij}$ can be
taken without loss of generality to lie in the first quadrant,
$\theta_{ij} \in [0, \pi/2]$, and the CP phase $\delta_\text{CP} \in
[0, 2\pi]$.
In the following we adopt the (arbitrary) convention of reporting
results for $\Dmq_{31}$ for NO and $\Dmq_{32}$ for IO, \textit{i.e.},
we always use the one which has the larger absolute value. Sometimes
we will generically denote such quantity as $\Dmq_{3\ell}$, with
$\ell=1$ for NO and $\ell=2$ for IO.

In this article, we present an up-to-date (as of summer 2014) global
analysis of solar, atmospheric, reactor and accelerator neutrino data
in the framework of three-neutrino oscillations. Alternative recent
global fits have been presented in Refs.~\cite{Capozzi:2013csa,
  Forero:2014bxa}.  In Sec.~\ref{sec:global} we describe the data used
in our analysis (listed also in Appendix~\ref{sec:appendix}) and we
present the results of the global analysis and the allowed ranges of
the oscillation parameters. In Sec.~\ref{sec:unitary} we focus on our
knowledge on CP violation, discussing the present status of the
leptonic Jarlskog invariant and displaying the results of our fit in
terms of leptonic unitarity triangles.  In Sec.~\ref{sec:tenten} we
comment on various ``tensions and tendencies'' in the global data,
including the reactor anomaly, the tension in the $\Dmq_{21}$
determination from solar experiments versus KamLAND, the determination
of $\Dmq_{31}$, tendencies in fit results for $\theta_{23}$ and
$\delta_\text{CP}$, and statistical issues related to the
determination of the CP violating phase $\delta_\text{CP}$. Finally in
Sec.~\ref{sec:summary} we present our conclusions.

The numerical results of our analysis as well as figures are available
at the website~\cite{nufit}, where also one- and two-dimensional $\chi^2$
tables are available for download. Furthermore, this website will be
kept up-to-date when new data becomes available.

%%%%%%%%%%%%%%%%%%%%%%%%%%%%%%%%%%%%%%%%%%%%%%%%%%%%%%%%%%%%%%%%%%%%%%%%%%%%%%

\section{Oscillation parameters: results of the global analysis}
\label{sec:global}

\subsection{Data included in our analysis}

We include in our global analysis the results from Super-Kamiokande
atmospheric neutrino data from phases SK1--4~\cite{skatm:nu2014},
adding the 1775 days of phase SK4 to their published results on phases
SK1--3~\cite{Wendell:2010md}. Concretely, we consider sub-GeV and
multi-GeV $e$-like and $\mu$-like fully contained events, as well as
partially contained, stopping and through-going $\mu$ data, each
divided into 10 angular bins. Hence we have a total of 70 energy and
zenith angle bins.
For what concerns disappearance results from long baseline accelerator
experiments (LBL) we use the energy distribution of events from MINOS
in both $\nu_\mu$ ($\bar\nu_\mu$) disappearance with $10.71 ~(3.36)
\times 10^{20}$ protons on target (pot)~\cite{Adamson:2013whj}, which
amounts to 39 (14) data points, and from T2K in $\nu_\mu$
disappearance~\cite{Abe:2014ugx} with $6.57\times 10^{20}$ pot (16 data
points).
For LBL appearance results we include both the neutrino and
antineutrino events from MINOS~\cite{Adamson:2013ue}, with exposure
$10.6 \times 10^{20}$ and $3.3 \times 10^{20}$ pot, respectively, and
from T2K in $\nu_e$ appearance~\cite{Abe:2013hdq} with $6.57\times
10^{20}$ pot; each of these samples contributes 5 data points.

In the analysis of solar neutrino experiments we include the total
rates from the radiochemical experiments
Chlorine~\cite{Cleveland:1998nv}, Gallex/GNO~\cite{Kaether:2010ag} and
SAGE~\cite{Abdurashitov:2009tn}. For real-time experiments we include
the results from on electron scattering (ES) from the four phases in
Super-Kamiokande: the 44 data points of the phase I (SK1)
energy-zenith spectrum~\cite{Hosaka:2005um}, the 33 (42) data points
of the full energy and day/night spectrum in phase II (III),
SK2~\cite{Cravens:2008aa} (SK3~\cite{Abe:2010hy}), and the 24 data
points of the energy spectrum and day-night asymmetry of the 1669-day
of phase IV, SK4~\cite{sksol:nu2014}.  The results of the three phases
of SNO are included in terms of the parametrization given in their
combined analysis~\cite{Aharmim:2011vm} which amount to 7 data points.
We also include the main set of the 740.7 days of Borexino
data~\cite{Bellini:2011rx} as well as their high-energy spectrum from
246 live days~\cite{Bellini:2008mr}.  In the analysis of solar
neutrino data we use the GS98 version of the solar standard
model~\cite{Serenelli:2009yc} (see Sec.~\ref{sec:tenten.dmq12}).

For oscillation signals at reactor experiments we include data from
the finalized experiments CHOOZ~\cite{Apollonio:1999ae} (energy
spectrum data, 14 data points) and Palo Verde~\cite{Piepke:2002ju}
(total rate) together with the spectrum from Double Chooz with 227.9
days live time~\cite{Abe:2012tg} (18 data points), and the 621-day
spectrum from Daya Bay~\cite{db:nu2014} (36 data points), as well as
the near and far rates observed at RENO with 800 days of
data-taking~\cite{reno:nu2014} (2 data points with free
normalization).  We also include the observed energy spectrum in
KamLAND data sets DS-1 and DS-2~\cite{Gando:2010aa} with a total
exposure of $3.49\times 10^{32}$ target-proton-year (2135 days).
Although reactor experiments with baselines $\lesssim 100$~m do not
contribute to oscillation physics, they play an important role in
constraining the unoscillated reactor neutrino flux. For this purpose
we consider also data from Bugey4~\cite{Declais:1994ma},
ROVNO4~\cite{Kuvshinnikov:1990ry}, Bugey3~\cite{Declais:1994su},
Krasnoyarsk~\cite{Vidyakin:1987ue, Vidyakin:1994ut},
ILL~\cite{Kwon:1981ua}, G\"osgen~\cite{Zacek:1986cu},
SRP~\cite{Greenwood:1996pb}, and ROVNO88~\cite{Afonin:1988gx}, to
which we refer as reactor short-baseline experiments (RSBL).  Details
on the RSBL analysis can be found in~\cite{Kopp:2013vaa}.

For convenience a detailed list of all the data used in our global
analysis can also be found in Appendix~\ref{sec:appendix}.

\subsection{Description of the results}
\label{sec:global.desc}

\begin{pagefigure}\centering
  \includegraphics[width=0.81\textwidth]{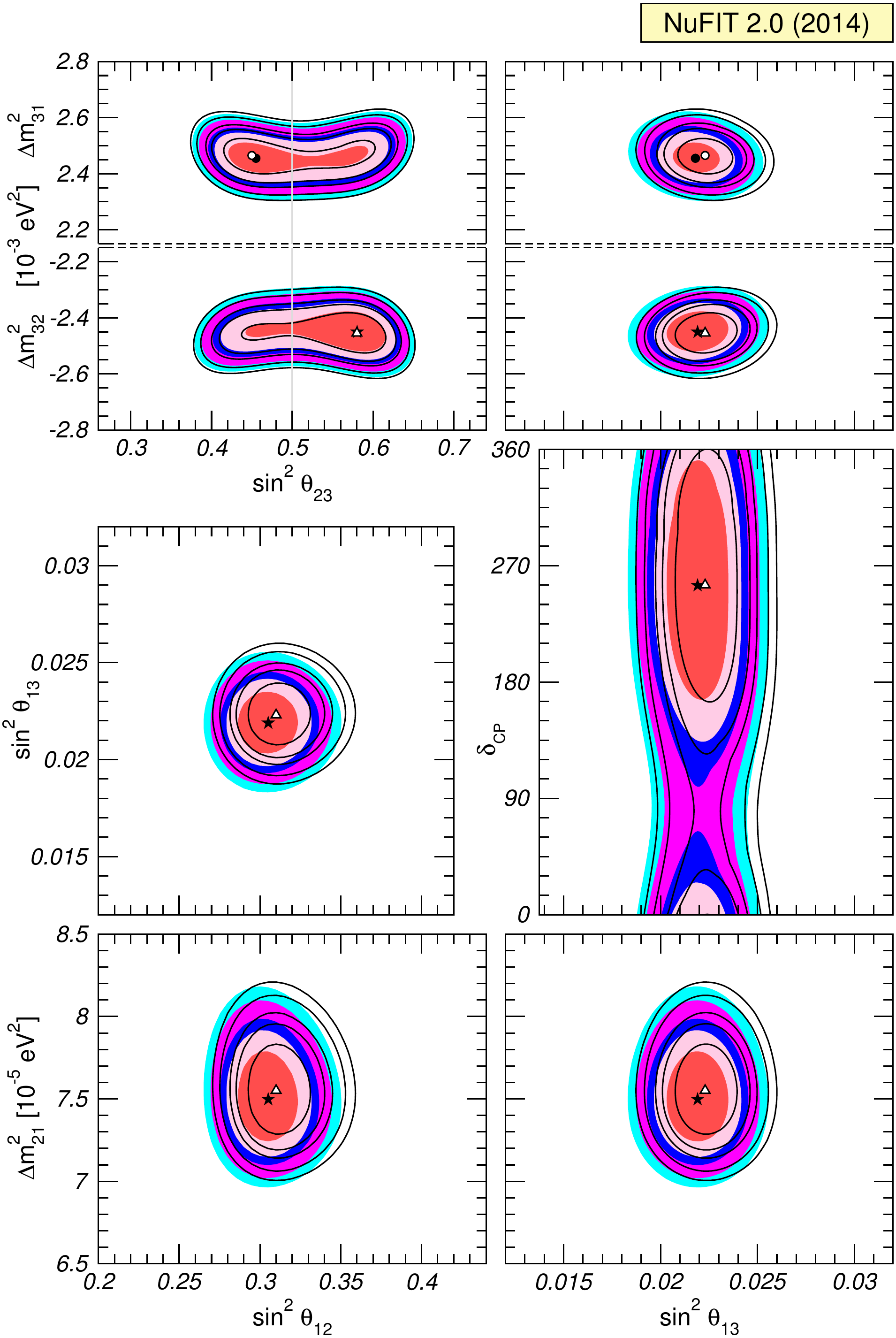}
  \caption{Global $3\nu$ oscillation analysis.  Each panel shows a
    two-dimensional projection of the allowed six-dimensional region
    after minimization with respect to the undisplayed parameters.
    The different contours correspond to $1\sigma$, 90\%, $2\sigma$,
    99\% and $3\sigma$ CL (2~dof). Full regions correspond to the
    analysis with free normalization of reactor fluxes and data from
    short-baseline (less than 100 m) reactor experiments included.
    For void regions short-baseline reactor data are not included but
    reactor fluxes as predicted in~\cite{Huber:2011wv} are
    assumed. Note that as atmospheric mass-squared splitting we use
    $\Dmq_{31}$ for NO and $\Dmq_{32}$ for IO. The regions in the
    lower 4 panels are based on a $\Delta\chi^2$ minimized with
    respect to NO and IO.}
  \label{fig:region-glob}
\end{pagefigure}

\begin{pagefigure}\centering
  \includegraphics[width=0.86\textwidth]{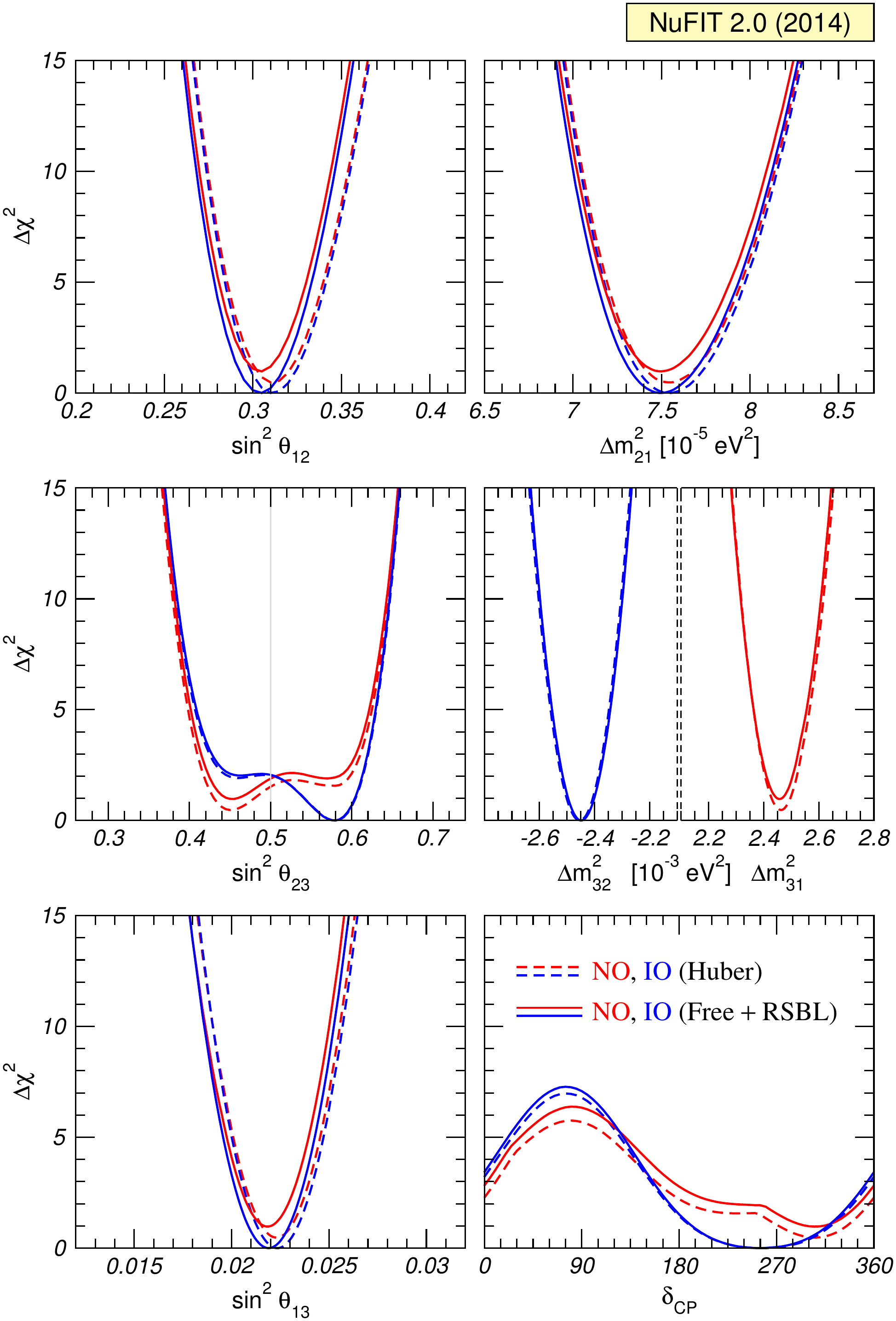}
  \caption{Global $3\nu$ oscillation analysis.  The red (blue) curves
    are for Normal (Inverted) Ordering.  For solid curves the
    normalization of reactor fluxes is left free and data from
    short-baseline (less than 100 m) reactor experiments are included.
    For dashed curves short-baseline data are not included but reactor
    fluxes as predicted in~\cite{Huber:2011wv} are assumed. Note that
    as atmospheric mass-squared splitting we use $\Dmq_{31}$ for NO
    and $\Dmq_{32}$ for IO.}
  \label{fig:chisq-glob}
\end{pagefigure}

The results of the global analysis are presented in
Figs.~\ref{fig:region-glob} and~\ref{fig:chisq-glob} where we show
different projections of the allowed six-dimensional parameter
space. To account for the possible effect of the so-called
\textit{reactor anomaly}~\cite{Mueller:2011nm, Huber:2011wv,
  Mention:2011rk}, we follow the approach of
Refs.~\cite{Schwetz:2011qt, GonzalezGarcia:2012sz} and study the
dependence of the determined value of the parameters on the
assumptions about the reactor fluxes.  To bracket the possible impact
of the anomaly, the results in Figs.~\ref{fig:region-glob}
and~\ref{fig:chisq-glob} are shown for two extreme choices.  The first
option is to leave the normalization of reactor fluxes free and
include data from short-baseline (less than 100 m) reactor
experiments. This corresponds to the colored regions in
Fig.~\ref{fig:region-glob} and the solid curves in
Fig.~\ref{fig:chisq-glob} (labeled ``Free+RSBL''). The second option
is not to include short-baseline reactor data but assume reactor
fluxes as predicted in~\cite{Huber:2011wv} (including their
uncertainties).  This corresponds to the black contours in
Fig.~\ref{fig:region-glob} and the dashed curves in
Fig.~\ref{fig:chisq-glob} (labeled ``Huber''). From the results in
these figures we conclude that:
\begin{enumerate}
\item for either choice of the reactor fluxes the global best fit
  corresponds to IO with $\sin^2\theta_{23} > 0.5$, while the second
  local minima is for NO and with $\sin^2\theta_{23} < 0.5$;

\item the statistical significance of the preference for Inverted
  versus Normal ordering is quite small, $\Delta\chi^2 \lesssim
  1\sigma$;

\item the present global analysis disfavors $\theta_{13}=0$ with a
  $\Delta\chi^2 \approx 500$. Such impressive result is mostly driven
  by the reactor data from Daya Bay with secondary contributions from
  RENO and Double Chooz;

\item the uncertainty on $\theta_{13}$ associated with the choice of
  reactor fluxes is reduced to the level of $0.5\sigma$ in the global
  analysis. This is so because the most precise results from Daya Bay
  and RENO are reactor flux normalization independent, as further
  discussed in Sec.~\ref{sec:tenten.flux};

\item a non-maximal value of the $\theta_{23}$ mixing is slightly
  favored, at the level of $\sim 1.4\sigma$ for Inverted Ordering at
  of $\sim 1.0\sigma$ for Normal Ordering;

\item the statistical significance of the preference of the fit for
  the second (first) octant of $\theta_{23}$ is $\leq 1.4\sigma$
  ($\leq 1.0\sigma$) for IO (NO);

\item the best fit for $\delta_\text{CP}$ for all analyses and
  orderings occurs for $\delta_\text{CP} \simeq 3\pi/2$, and values
  around $\pi/2$ are disfavored with $\Delta\chi^2 \simeq 6$.  A
  discussion on the corresponding CL can be found in
  Sec.~\ref{sec:tenten.stat}.
\end{enumerate}

\begin{table}\centering
  \begin{footnotesize}
    \begin{tabular}{l|cc|cc|c}
      \hline\hline
      & \multicolumn{2}{c|}{Normal Ordering ($\Delta\chi^2=0.97$)}
      & \multicolumn{2}{c|}{Inverted Ordering (best fit)}
      & Any Ordering
      \\
      \hline
      & bfp $\pm 1\sigma$ & $3\sigma$ range
      & bfp $\pm 1\sigma$ & $3\sigma$ range
      & $3\sigma$ range
      \\
      \hline
      \rule{0pt}{4mm}\ignorespaces
      $\sin^2\theta_{12}$
      & $0.304_{-0.012}^{+0.013}$ & $0.270 \to 0.344$
      & $0.304_{-0.012}^{+0.013}$ & $0.270 \to 0.344$
      & $0.270 \to 0.344$
      \\[1mm]
      $\theta_{12}/^\circ$
      & $33.48_{-0.75}^{+0.78}$ & $31.29 \to 35.91$
      & $33.48_{-0.75}^{+0.78}$ & $31.29 \to 35.91$
      & $31.29 \to 35.91$
      \\[3mm]
      $\sin^2\theta_{23}$
      & $0.452_{-0.028}^{+0.052}$ & $0.382 \to 0.643$
      & $0.579_{-0.037}^{+0.025}$ & $0.389 \to 0.644$
      & $0.385 \to 0.644$
      \\[1mm]
      $\theta_{23}/^\circ$
      & $42.3_{-1.6}^{+3.0}$ & $38.2 \to 53.3$
      & $49.5_{-2.2}^{+1.5}$ & $38.6 \to 53.3$
      & $38.3 \to 53.3$
      \\[3mm]
      $\sin^2\theta_{13}$
      & $0.0218_{-0.0010}^{+0.0010}$ & $0.0186 \to 0.0250$
      & $0.0219_{-0.0010}^{+0.0011}$ & $0.0188 \to 0.0251$
      & $0.0188 \to 0.0251$
      \\[1mm]
      $\theta_{13}/^\circ$
      & $8.50_{-0.21}^{+0.20}$ & $7.85 \to 9.10$
      & $8.51_{-0.21}^{+0.20}$ & $7.87 \to 9.11$
      & $7.87 \to 9.11$
      \\[3mm]
      $\delta_\text{CP}/^\circ$
      & $306_{-70}^{+39}$ & $\hphantom{00}0 \to 360$
      & $254_{-62}^{+63}$ & $\hphantom{00}0 \to 360$
      & $\hphantom{00}0 \to 360$
      \\[3mm]
      $\dfrac{\Dmq_{21}}{10^{-5}~\eVq}$
      & $7.50_{-0.17}^{+0.19}$ & $7.02 \to 8.09$
      & $7.50_{-0.17}^{+0.19}$ & $7.02 \to 8.09$
      & $7.02 \to 8.09$
      \\[3mm]
      $\dfrac{\Dmq_{3\ell}}{10^{-3}~\eVq}$
      & $+2.457_{-0.047}^{+0.047}$ & $+2.317 \to +2.607$
      & $-2.449_{-0.047}^{+0.048}$ & $-2.590 \to -2.307$
      & $\begin{bmatrix}
        +2.325 \to +2.599\\[-2pt]
        -2.590 \to -2.307
      \end{bmatrix}$
      \\[3mm]
      \hline\hline
    \end{tabular}
  \end{footnotesize}
  \caption{Three-flavor oscillation parameters from our fit to global
    data after the NOW~2014 conference. The results are presented for
    the ``Free Fluxes + RSBL'' in which reactor fluxes have been left
    free in the fit and short baseline reactor data (RSBL) with $L
    \lesssim 100$~m are included. The numbers in the 1st (2nd) column
    are obtained assuming NO (IO), \textit{i.e.}, relative to the
    respective local minimum, whereas in the 3rd column we minimize
    also with respect to the ordering. Note that $\Dmq_{3\ell} \equiv
    \Dmq_{31} > 0$ for NO and $\Dmq_{3\ell} \equiv \Dmq_{32} < 0$ for
    IO.}
  \label{tab:results}
\end{table}

In what follows we will consider our default analysis choice the one
with ``Free Fluxes + RSBL''. It is for this choice of fluxes that the
best fit values and the derived ranges for the six parameters at the
$1\sigma$ ($3\sigma$) level are given in Tab.~\ref{tab:results}.  For
each parameter the ranges are obtained after marginalizing with
respect to the other parameters.
We show the results for three scenarios.  In the first and second
columns we assume that the ordering of the neutrino mass states is
known ``a priori'' to be Normal or Inverted, respectively, so the
ranges of all parameters are defined with respect to the minimum in
the given scenario.
In the third column we make no assumptions on the ordering, so in this
case the ranges of the parameters are defined with respect to the
global minimum (which corresponds to Inverted Ordering) and are
obtained marginalizing also over the ordering. For this third case we
only give the $3\sigma$ ranges. Of course in this case the range of
$\Dmq_{3\ell}$ is composed of two disconnected intervals, one one
containing the absolute minimum (IO) and the other the secondary local
minimum (NO).

Let us define the $3\sigma$ relative precision of a parameter by
$2(x^\text{up} - x^\text{low}) / (x^\text{up} + x^\text{low})$, where
$x^\text{up}$ ($x^\text{low}$) is the upper (lower) bound on a
parameter $x$ at the $3\sigma$ level. From the numbers in the table we
then find $3\sigma$ relative precisions of 14\% ($\theta_{12}$), 32\%
($\theta_{23}$), 15\% ($\theta_{13}$), 14\% ($\Dmq_{21}$) and 11\%
($|\Dmq_{3\ell}|$) for the various oscillation parameters.

\section{Mixing matrix and leptonic CP violation}
\label{sec:unitary}

From the global $\chi^2$ analysis described in the previous section
and following the procedure outlined in
Ref.~\cite{GonzalezGarcia:2003qf} one can derive the $3\sigma$ ranges
on the magnitude of the elements of the leptonic mixing matrix to be:
\begin{equation}
  \label{eq:umatrix}
  |U| = \begin{pmatrix}
    0.801 \to 0.845 &\qquad
    0.514 \to 0.580 &\qquad
    0.137 \to 0.158
    \\
    0.225 \to 0.517 &\qquad
    0.441 \to 0.699 &\qquad
    0.614 \to 0.793
    \\
    0.246 \to 0.529 &\qquad
    0.464 \to 0.713 &\qquad
    0.590 \to 0.776
  \end{pmatrix} .
\end{equation}
By construction the derived limits in Eq.~\eqref{eq:umatrix} are
obtained under the assumption of the matrix $U$ being unitary.  In
other words, the ranges in the different entries of the matrix are
correlated due to the constraints imposed by unitarity, as well as the
fact that, in general, the result of a given experiment restricts a
combination of several entries of the matrix.  As a consequence
choosing a specific value for one element further restricts the range
of the others.

\begin{figure}\centering
  \includegraphics[width=0.9\textwidth]{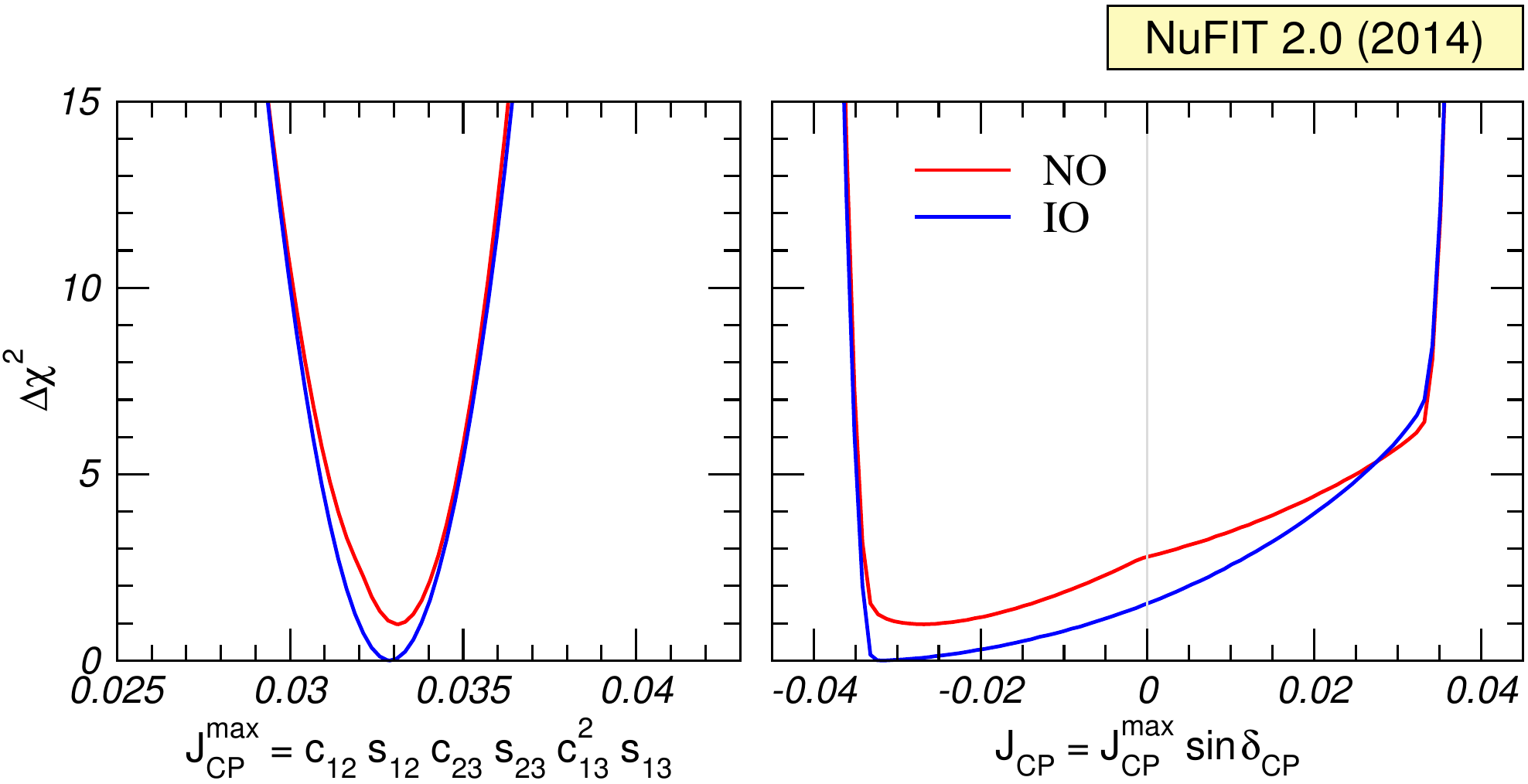}
  \caption{Dependence of the global $\Delta\chi^2$ function on the
    Jarlskog invariant. The red (blue) curves are for NO (IO).}
  \label{fig-chisq-viola}
\end{figure}

The present status of the determination of leptonic CP violation is
illustrated in Fig.~\ref{fig-chisq-viola} where we show the dependence
of the $\Delta\chi^2$ of the global analysis on the Jarlskog invariant
which gives a convention-independent measure of CP
violation~\cite{Jarlskog:1985ht}, defined as usual by:
\begin{equation}
  \Im\big[ U_{\alpha i} U_{\alpha j}^* U_{\beta i}^* U_{\beta j} \big]
  \equiv \sum_{\gamma=e,\mu,\tau} \sum_{k=1,2,3} \,
  J_\text{CP} \, \epsilon_{\alpha\beta\gamma} \, \epsilon_{ijk}
  \equiv J_\text{CP}^\text{max} \, \sin\delta_\text{CP} \,.
\end{equation}
Using the parametrization in Eq.~\eqref{eq:matrix} we get
\begin{equation}
  J^\text{max}_\text{CP} = \cos\theta_{12} \sin\theta_{12}
  \cos\theta_{23} \sin\theta_{23} \cos^2\theta_{13} \sin\theta_{13} \,.
\end{equation}
From the left panel of Fig.~\ref{fig-chisq-viola} we see that the
determination of the mixing angles yields at present a maximum allowed
CP violation
\begin{equation}
  \label{eq:jmax}
  J_\text{CP}^\text{max} = 0.0329 \pm 0.0009 \; (\mathrel{\pm} 0.0027)
\end{equation}
at $1\sigma$ ($3\sigma$) for both orderings. The preference of the
present data for non-zero $\delta_\text{CP}$ implies a best fit
$J_\text{CP}^\text{best} = -0.032$, which is favored over CP
conservation at the $\sim 1.2\sigma$ level. These numbers can be
compared with the size of the Jarlskog invariant in the quark sector,
which is determined to be $J_\text{CP}^\text{quarks} =
(2.96^{+0.20}_{-0.16}) \times 10^{-5}$~\cite{PDG}.

\begin{figure}\centering
  \includegraphics[width=0.9\textwidth]{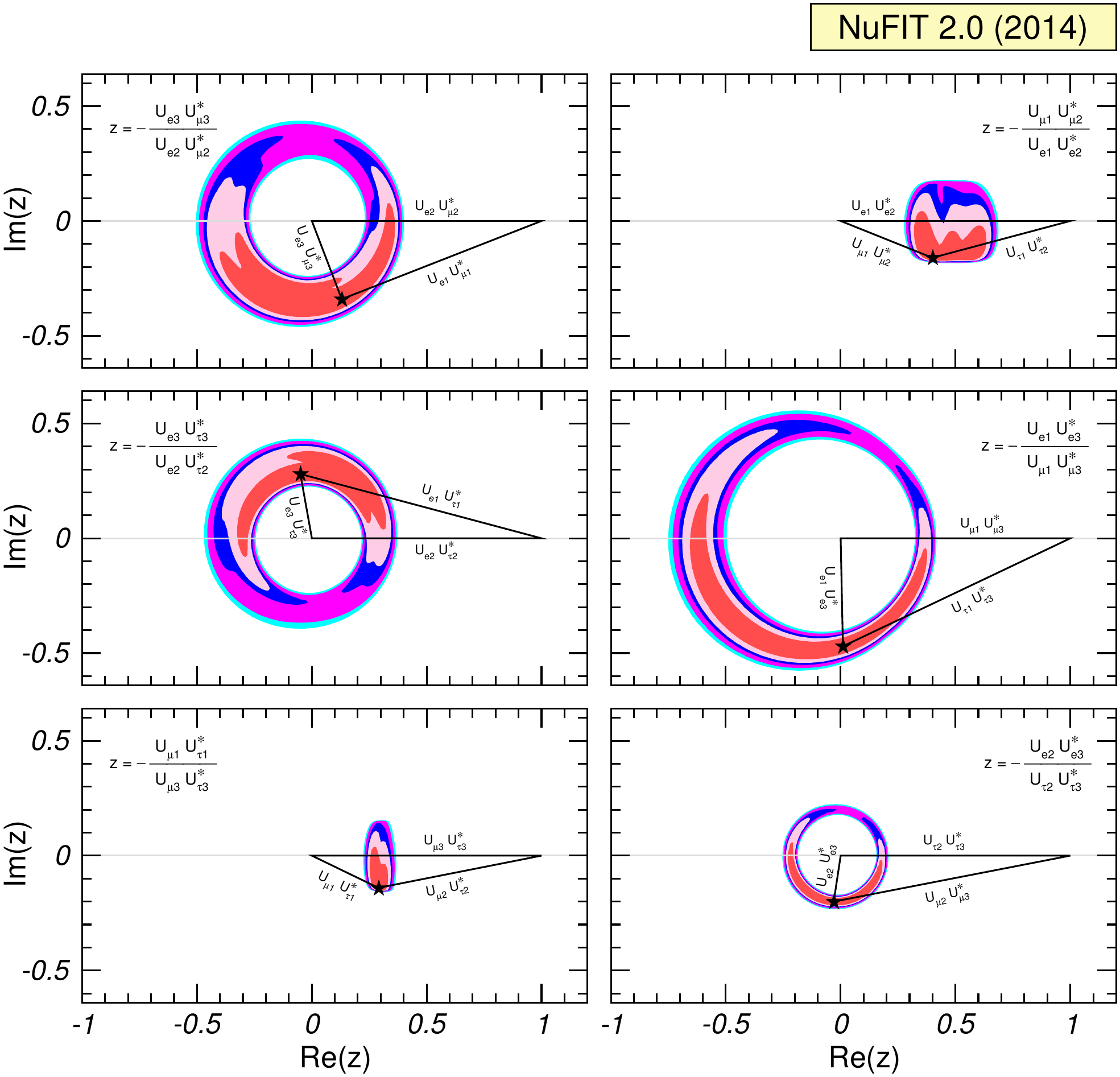}
  \caption{Six leptonic unitarity triangles. After scaling and
    rotating each triangle so that two of its vertices always coincide
    with $(0,0)$ and $(1,0)$ (see text for details) we plot the
    $1\sigma$, 90\%, $2\sigma$, 99\%, $3\sigma$ CL (2~dof) allowed
    regions of the third vertex. Note that in the construction of the
    triangles the unitarity of the $U$ matrix is always explicitly
    imposed.}
  \label{fig:region-viola}
\end{figure}

In Fig.~\ref{fig:region-viola} we recast the allowed regions for the
leptonic mixing matrix in terms of leptonic unitarity triangles, which
are obtained as different combinations of the entries of the $U$
matrix.\footnote{See, \textit{e.g.}, Refs.~\cite{Farzan:2002ct,
    Smirnov:2008nh, Dueck:2010fa, He:2013rba} for discussions of
  leptonic unitarity triangles.} Since in our analysis $U$ is unitary
by construction, any given pair of rows or columns can be used to
define a triangle in the complex plane. On the left (right) panels we
show the triangles corresponding to the unitarity conditions
\begin{equation}
  \begin{aligned}
    \sum_{i=1,2,3} U_{\alpha  i}U^*_{\beta i} &= 0
    && \text{with~} \alpha \neq \beta
    && \text{(left),}
    \\
    \sum_{\alpha=e,\mu,\tau} U_{\alpha  i}U^*_{\alpha j} &= 0
    && \text{with~} i \neq j
    && \text{(right).}
  \end{aligned}
\end{equation}
In drawing these triangles we have rescaled and rotated their sides so
that two of their vertices always coincide with $(0,0)$ and $(1,0)$ in
the complex plane. To this aim we have defined a complex variable $z$
as follows:
\begin{equation}
  \begin{aligned}
    z = -\frac{U_{\alpha i} U^*_{\beta i}}{U_{\alpha k} U^*_{\beta k}}
    &= 1 + \frac{U_{\alpha j} U^*_{\beta  j}}{U_{\alpha k}U^*_{\beta k}}
    && \text{with~} \alpha \neq \beta \text{~and~} i \neq j \neq k
    && \text{(left),}
    \\
    z = -\frac{U_{\alpha i} U^*_{\alpha j}}{U_{\gamma i} U^*_{\gamma j}}
    &= 1 + \frac{U_{\beta i} U^*_{\beta j}}{U_{\gamma i}U^*_{\gamma j}}
    && \text{with~} i \neq j \text{~and~} \alpha \neq \beta \neq \gamma
    && \text{(right)}
  \end{aligned}
\end{equation}
and then we have plot the $1\sigma$, 90\%, $2\sigma$, 99\%, $3\sigma$
CL (2~dof) allowed regions of the third vertex of the triangle as the
real and imaginary parts of $z$.  For convenience in each panel we
have chosen the \emph{normalization side} (the one which lies on the
horizontal $(0,0) \to (0,1)$ segment) as the best determined of the
two longer sides of each triangle. In this way all the triangles have
more or less the same size, and the uncertainty in the position of the
third vertex is not too much affected by the uncertainty of the
normalization side.
Note that the most common unitarity triangle in the quark sector is
the one based on the $d$-quark and $b$-quark columns~\cite{PDG}, which
corresponds to the 1st and 3rd column in the leptonic matrix,
\textit{i.e.}, to the triangle in the middle-right panel in
Fig.~\ref{fig:region-viola}.

In this kind of diagrams the absence of CP violation implies a flat
triangle, \textit{i.e.}, $\Im(z) = 0$.  As can be seen, in all the
panels the horizontal axis marginally crosses the $1\sigma$ allowed
region, which for 2~dof corresponds to $\Delta\chi^2 \simeq 2.3$. This
is consistent with the present preference for CP violation,
$\chi^2(J_\text{CP} = 0) - \chi^2(J_\text{CP}~\text{free}) = 1.5$.
%

%%%%%%%%%%%%%%%%%%%%%%%%%%%%%%%%%%%%%%%%%%%%%%%%%%%%%%%%%%%%%%%%%%%%%%%%%%%%%%%

\section{Tension and tendencies}
\label{sec:tenten}

\subsection{Impact of reactor flux uncertainties}
\label{sec:tenten.flux}

\begin{figure}\centering
  \includegraphics[width=0.5\textwidth]{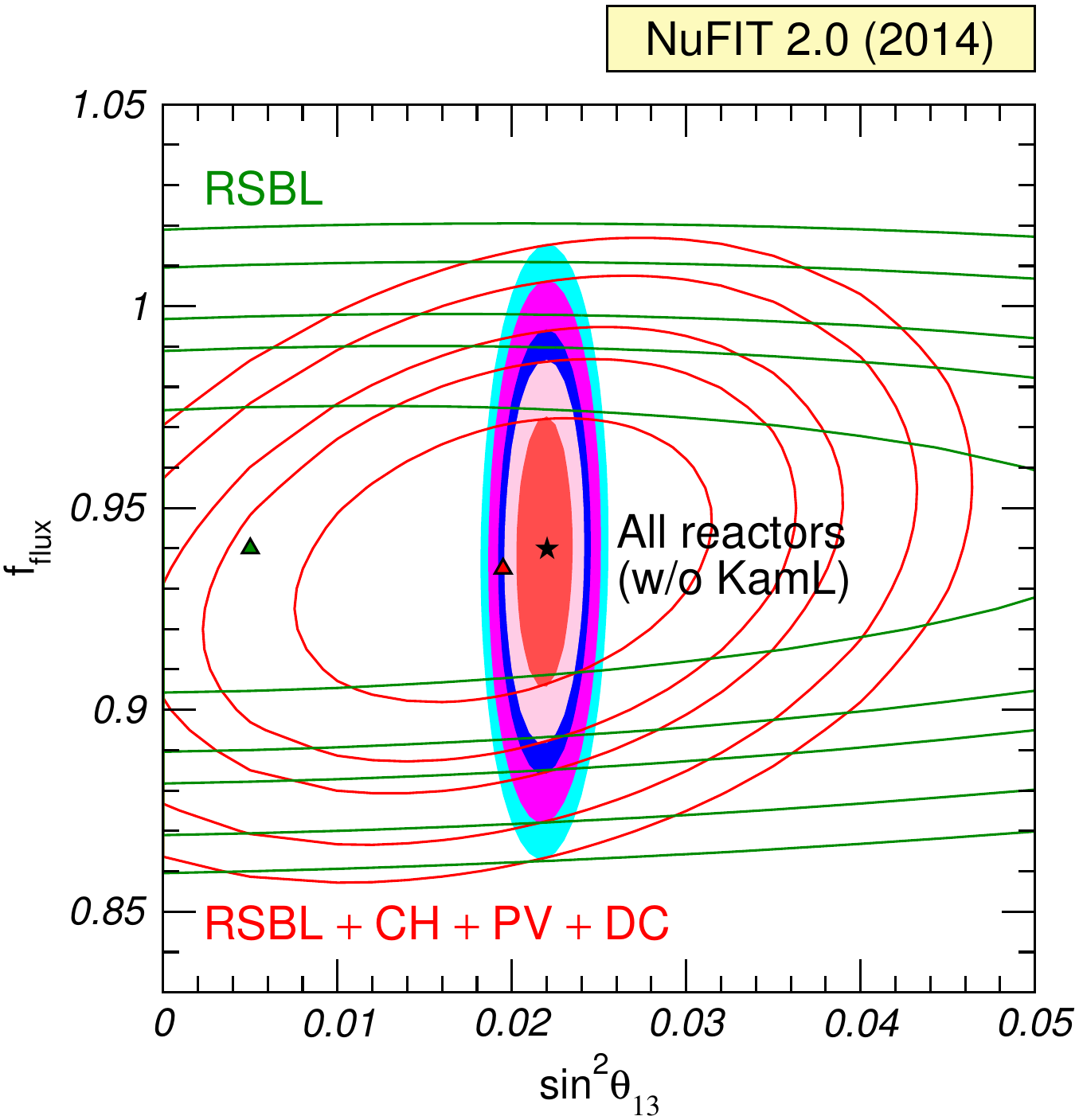}
  \caption{Contours ($1\sigma$, 90\%, $2\sigma$, 99\%, $3\sigma$ CL
    for 2~dof) in the plane of $\theta_{13}$ and the reactor flux
    normalization $f_\text{flux}$.  Full regions correspond to the
    combined analysis of all reactor neutrino experiments with the
    exception of KamLAND, but including the RSBL experiments.  The
    green contours correspond to only the RSBL experiments and red
    contours include RSBL + medium-baseline reactors without a near
    detector (\textit{i.e.} without including Daya Bay and RENO).}
  \label{fig:react-t13}
\end{figure}

Within the 3-flavor framework the so-called \textit{reactor anomaly}
leads to a ``tension'' of about $2.7\sigma$ between the predicted
reactor neutrino fluxes~\cite{Mueller:2011nm, Huber:2011wv} and the
event rates observed in short-baseline reactor experiments.  By
adopting two extreme approaches in dealing with this tension we have
shown in Sec.~\ref{sec:global.desc} that the impact on the
determination of the oscillation parameters in the global fit is quite
small, at the level of $0.5\sigma$ for $\sin^2\theta_{13}$ (see
Figs.~\ref{fig:region-glob} and~\ref{fig:chisq-glob}).
This is further illustrated in Fig.~\ref{fig:react-t13} where we show
the allowed regions in the plane of $\theta_{13}$ and the flux
normalization $f_\text{flux}$ (relative to the one predicted
in~\cite{Huber:2011wv}) for several combinations of the reactor
experiments. Short-baseline data (green contours) essentially
determine the flux normalization. Adding also data from experiments at
around 1~km without a dedicated near detector (red contours) provides
already a signal for non-zero $\theta_{13}$, but such result is
affected by significant correlation with the flux
normalization. However, once the precise data on near-far comparison
from Daya Bay and RENO are included (colored regions) no correlation
is left between the determination of $\theta_{13}$ and
$f_\text{flux}$. Thus in the $3\nu$ analysis the unexplained reactor
anomaly mostly translates in an overall increase of the $\chi^2$ in
the analysis with fluxes from Ref.~\cite{Huber:2011wv} with
$\chi^2(f_\text{flux} = 1) - \chi^2(f_\text{flux}~\text{free}) \simeq
7$.  Details of our analysis in this respect can be found in
Ref.~\cite{Kopp:2013vaa}, where a discussion of a possible explanation
in terms of sterile neutrinos is also given.

\subsection{Determination of $\Dmq_{21}$: solar and KamLAND}
\label{sec:tenten.dmq12}

\begin{figure}\centering
  \includegraphics[width=0.9\textwidth]{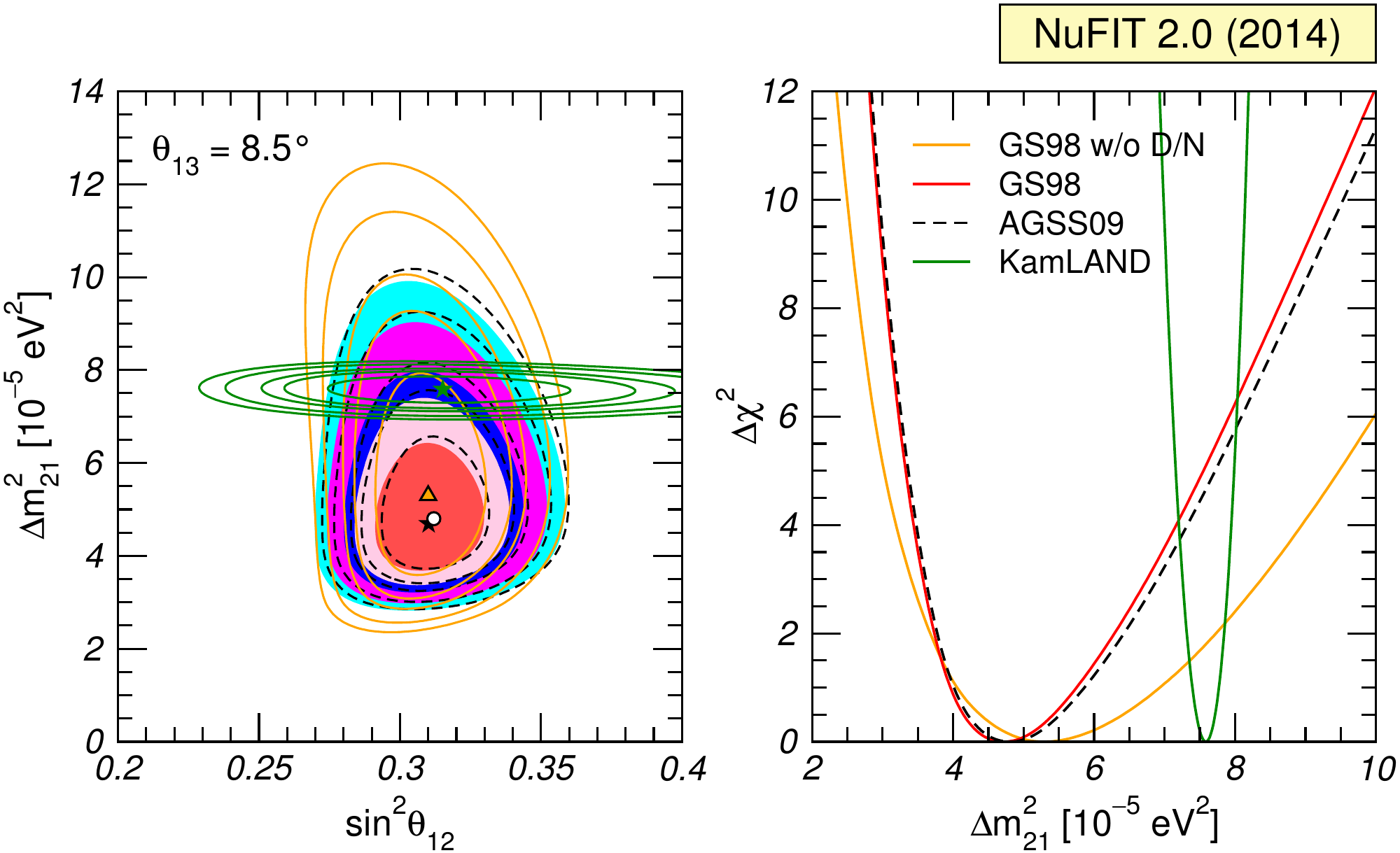}
  \caption{Left: Allowed parameter regions (at $1\sigma$, 90\%,
    $2\sigma$, 99\% and $3\sigma$ CL for 2~dof) from the combined
    analysis of solar data for GS98 model (full regions with best fit
    marked by black star) and AGSS09 model (dashed void contours with
    best fit marked by a white dot), and for the analysis of KamLAND
    data (solid green contours with best fit marked by a green star)
    for fixed $\theta_{13}=8.5^\circ$. We also show as orange contours
    the results of a global analysis for the GS98 model but without
    including the day-night information from SK (see text for
    details).  Right: $\Delta\chi^2$ dependence on $\Dmq_{21}$ for the
    same four analysis after marginalizing over $\theta_{12}$.}
  \label{fig:sun-tension}
\end{figure}

We show in Fig.~\ref{fig:sun-tension} the results of the analysis of
the solar experiments and of KamLAND which give the dominant
contribution to the determination of $\Dmq_{21}$ and $\theta_{12}$.
Here $\theta_{13}$ is fixed to the present best fit value of the
global analysis.  For the sake of completeness the solar neutrino
results are shown for two different versions of the Standard Solar
Model, namely the GS98 and the AGSS09
models~\cite{Serenelli:2009yc}. Let us remind that GS98 is based on
the older solar abundances leading to high metallicity and which
perfectly agreed with helioseismological data, whereas AGSS09 uses the
new precise determination of the solar abundances which imply a lower
metallicity and cannot reproduce the helioseismological data. This
conflict constitutes the so-called ``solar composition
problem''. Although it is a pretty serious problem in the context of
solar physics, its impact in the determination of the relevant
oscillation parameters is very small, as can be seen clearly from
Fig.~\ref{fig:sun-tension}.

The left panel in Fig.~\ref{fig:sun-tension} illustrates the
complementarity of solar and KamLAND in the determination of the
``12'' parameters.  Solar experiments provide the best precision of
$\theta_{12}$ while KamLAND gives a better determination of
$\Dmq_{21}$. We remind the reader that the relevant survival
probabilities for these experiments in the framework of three neutrino
oscillations can be written as:
\begin{equation}
  \label{eq:ps3}
  P^{3\nu}_{ee} = \sin^4\theta_{13} + \cos^4\theta_{13}
  P^{2\nu}_{ee}(\Dmq_{21},\theta_{12}) \,,
\end{equation}
where we have used the fact that $L^\text{osc}_{31} = 4\pi E_\nu /
\Dmq_{31}$ is much shorter than the distance traveled by both solar
and KamLAND neutrinos, so that the oscillations related to
${L_{31}^\text{osc}}$ are averaged.  In presence of matter effects
$P^{2\nu}_{ee}(\Dmq_{21},\theta_{12})$ should be calculated taking
into account the evolution in an effective matter density
$n^\text{eff}_{e} = n_e \cos^2\theta_{13}$.  For $10^{-5}\lesssim
\Dmq/\eVq \lesssim 10^{-4}$, $P^{2\nu}_{ee}(\Dmq_{21},\theta_{12})$
presents the following asymptotic behaviors~\cite{Goswami:2004cn}:
\begin{align}
  \label{eq:ps2l}
  P^{2\nu,\text{sun}}_{ee}
  & \simeq 1 - \frac{1}{2} \sin^2(2\theta_{12})
  & \text{for~} E_\nu
  & \lesssim \text{few} \times 100~\text{KeV}
  \\
  \label{eq:ps2h}
  P^{2\nu,\text{sun}}_{ee}
  & \simeq \sin^2(\theta_{12})
  & \text{for~} E_\nu
  & \gtrsim \text{few} \times 1~\text{MeV}
  \\
  \label{eq:2kam}
  P_{ee}^{2\nu, \text{kam}}
  &= 1 - \frac{1}{2}\sin^2(2\theta_{12}) \sin^2\frac{\Dmq_{21} L}{2 E_\nu} \,.
\end{align}
At present most of the precision of the solar analysis is provided by
SNO and SK for which the relevant MSW survival
probability~\cite{Wolfenstein:1977ue, Mikheev:1986gs} provides a
direct measurement of $\sin^2\theta_{12}$, as seen in
Eq.~\eqref{eq:ps2h}.  In the MSW regime the determination of
$\Dmq_{21}$ in solar experiments comes dominantly from the ratio
between the solar potential and the $\Dmq_{21}$ term required to
simultaneously describe the CC/NC data at SNO and the undistorted
spectra of \Nuc{8}{B} neutrinos as measured in both SK and SNO.
Conversely KamLAND $\bar\nu_e$ survival probability proceeds
dominantly as vacuum oscillations and provides a most precise
determination of $\Dmq_{21}$ via the strong effect of the oscillating
phase in the distortion of the reactor energy spectrum. On the
contrary it yields a weaker constraint on $\theta_{12}$ as the vacuum
oscillation probability depends on the double-valued and ``flatter''
function $\sin^2(2\theta_{12})$.

As seen in the left panel in Fig.~\ref{fig:sun-tension} for either
version of the solar model the best fit points of solar and KamLAND
analysis lie at very similar values of $\theta_{12}$.  As it was
pointed out in Ref.~\cite{Fogli:2008jx} and widely discussed in the
literature~\cite{Fogli:2009zza, Schwetz:2008er, Maltoni:2008ka,
  Balantekin:2008zm, GonzalezGarcia:2010er}, the matching in the
determination of $\theta_{12}$ requires the presence of a non-zero
value of $\theta_{13}$.  With the present determination of
$\theta_{13}$ provided by the medium baseline reactor experiments, the
agreement between the best fit point values of $\theta_{12}$ is
remarkable.

From the same figure, however, we see that the value of $\Dmq_{21}$
preferred by KamLAND is higher than the one from solar experiments.
At present this is about a $2\sigma$ effect, as can be seen in the
right panel where we show the $\Delta\chi^2$ dependence as a function
of $\Dmq_{21}$ when marginalized over $\theta_{12}$. 
This tension has been present during the last two years and it arises
from a combination of two effects: (a) the well-known fact that none
of the \Nuc{8}{B} measurement performed by SNO, SK and Borexino show
any evidence of the spectrum low energy turn-up expected in the
standard LMA-MSW solution, and (b) the indication of a non-vanishing
day-night asymmetry in SK, which disfavors the KamLAND $\Dmq_{21}$
best fit value for which Earth matter effects are too small. The
relevance of these effects is illustrated in
Fig.~\ref{fig:sun-tension} where we show the results of our analysis
both with and without the inclusion of the SK day-night information.
As can be seen, once the SK day-night information is removed the solar
best-fit point shifts upwards and the solar allowed region extends to
much larger values of $\Dmq_{21}$, as expected, so that the tension
with KamLAND is reduced to about $1.4\sigma$.
Modified matter potential due to non-standard
interactions~\cite{Palazzo:2011vg, Gonzalez-Garcia:2013usa} and
super-light sterile neutrinos~\cite{deHolanda:2010am} have been
proposed as extended scenarios which could relax this tension.

\subsection{Determination of $\Dmq_{3\ell}$: $\nu_\mu$ and $\nu_e$ disappearance}
\label{sec:tenten.dmq3l}

\begin{figure}\centering
  \includegraphics[width=0.9\textwidth]{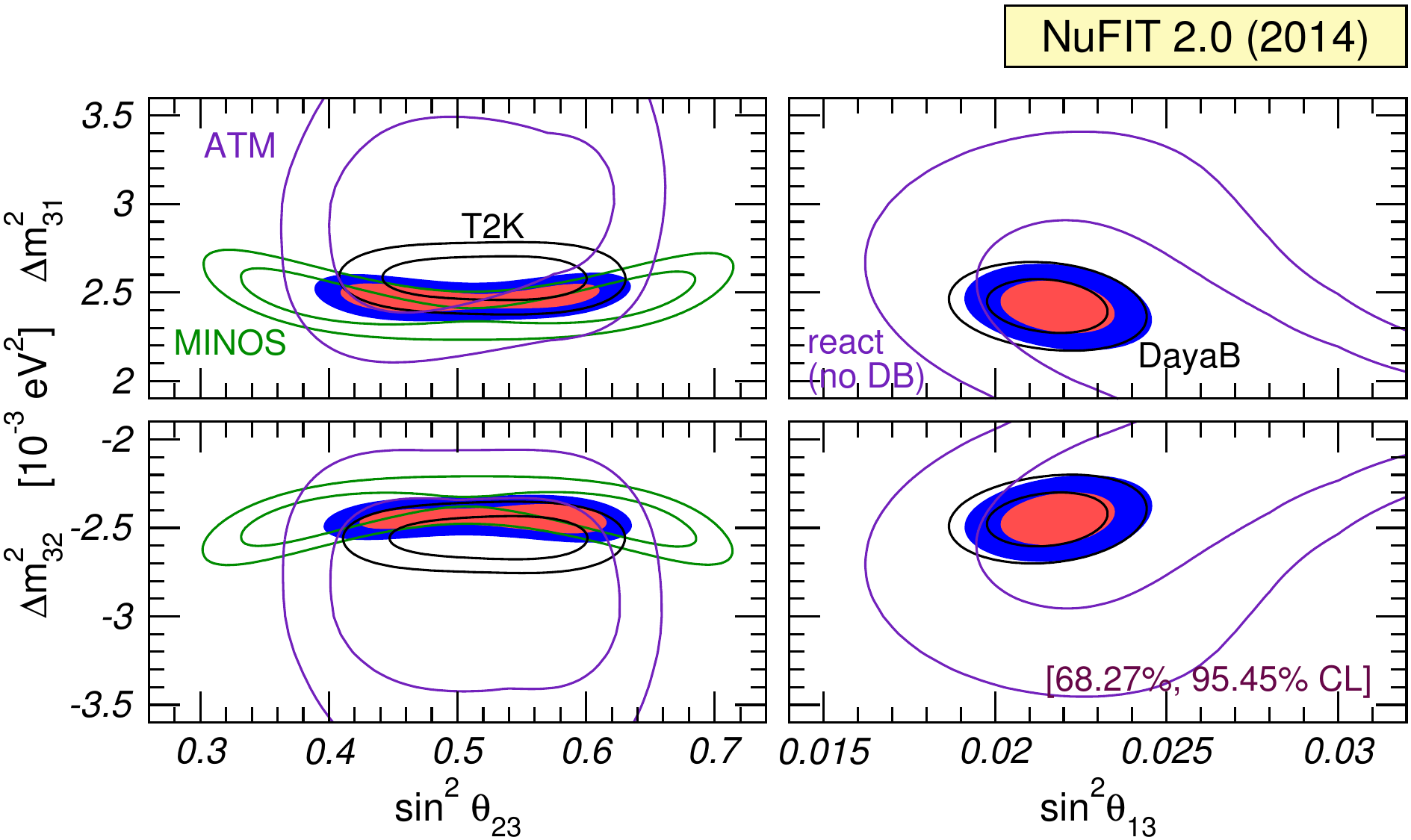}
  \caption{Determination of $\Dmq_{3\ell}$ at $1\sigma$ and $2\sigma$
    (2~dof), where $\ell=1$ for NO (upper panels) and $\ell=2$ for IO
    (lower panels). The left panels show regions in the
    $(\sin^2\theta_{23}, \Dmq_{3\ell})$ plane using both appearance
    and disappearance data from MINOS (green) and T2K (black), as well
    as SK atmospheric data (green) and a combination of them (colored
    regions). Here $\theta_{13}$ is constrained to the $3\sigma$ range
    from the global fit.  The right panels show regions in the
    $(\sin^2\theta_{13}, \Dmq_{3\ell})$ plane using data from Daya Bay
    (black), reactor data without Daya Bay (violet), and their
    combination (colored regions). In all panels solar and KamLAND
    data are included to constrain $\Dmq_{21}$ and
    $\theta_{12}$. Contours are defined with respect to the local
    minimum in each panel.}
  \label{fig:dmq3l-samples}
\end{figure}

Fig.~\ref{fig:dmq3l-samples} illustrates the determination of
$\Dmq_{3\ell}$ from different data sets. In the left panels we focus
on long-baseline $\nu_\mu$ disappearance data. It is clear that in
this case the final precision on $|\Dmq_{3\ell}|$ emerges from the
combination of T2K and MINOS data, while the determination of
$\sin^2\theta_{23}$ is dominated by T2K.

Concerning $\nu_e$ disappearance data, Eq.~\eqref{eq:peereac} in
Sec.~\ref{sec:tenten.sets} implies that the rates observed in reactor
experiments at different baselines can provide an independent
determination of $\Dmq_{3\ell}$~\cite{Bezerra:2012at,
  GonzalezGarcia:2012sz}. On top of this, the observation of the
energy-dependent oscillation effect of $\theta_{13}$ in Daya
Bay~\cite{An:2013zwz} allows a rather precise determination of
$|\Dmq_{3\ell}|$.  In the right panels of Fig.~\ref{fig:dmq3l-samples}
we show therefore the allowed regions in the $(\theta_{13},
\Dmq_{3\ell})$ plane based on global data on $\nu_e$
disappearance. The blue contours are obtained from all the
medium-baselines reactor experiments with the exception of Daya
Bay. Those regions emerge from the baseline effect mentioned
above. The black contour are based on the energy spectrum in Daya Bay,
whereas the colored regions show the combination.

By comparing the left and right panels we observe that $\nu_\mu$ and
$\nu_e$ disappearance experiments by now provide a consistent
determination of $|\Dmq_{3\ell}|$ with similar precision.

\subsection{Mass ordering, $\theta_{23}$ octant and CP phase: role of different data sets}
\label{sec:tenten.sets}

As we have seen in Sec.~\ref{sec:global}, several 1-2$\sigma$
``tendencies'' appear in the global analysis in the determination of
the mass ordering, the octant of $\theta_{23}$, and the CP violating
phase. To illustrate the role of the different data sets on such
tendencies, we show in Fig.~\ref{fig:chisq-hier} the $\Delta\chi^2$ as
a function of $\Dmq_{3\ell}$, $\theta_{23}$, and $\delta_\text{CP}$
for different combinations of experiments.  In each panel the results
have been marginalized with respect to all undisplayed parameters
\emph{except} the mass ordering, which is fixed to Inverted (Normal)
for the left (right) panels. Note, however, that for each combination
of experiments the $\Delta\chi^2$ is defined with respect to the
absolute minima between the two orderings. In this way the difference
between the ``height'' of the minimum of the curve on the left and the
corresponding one on the right gives the contribution of that set of
observables to the determination of the mass ordering.

\begin{pagefigure}\centering
  \includegraphics[width=0.9\textwidth]{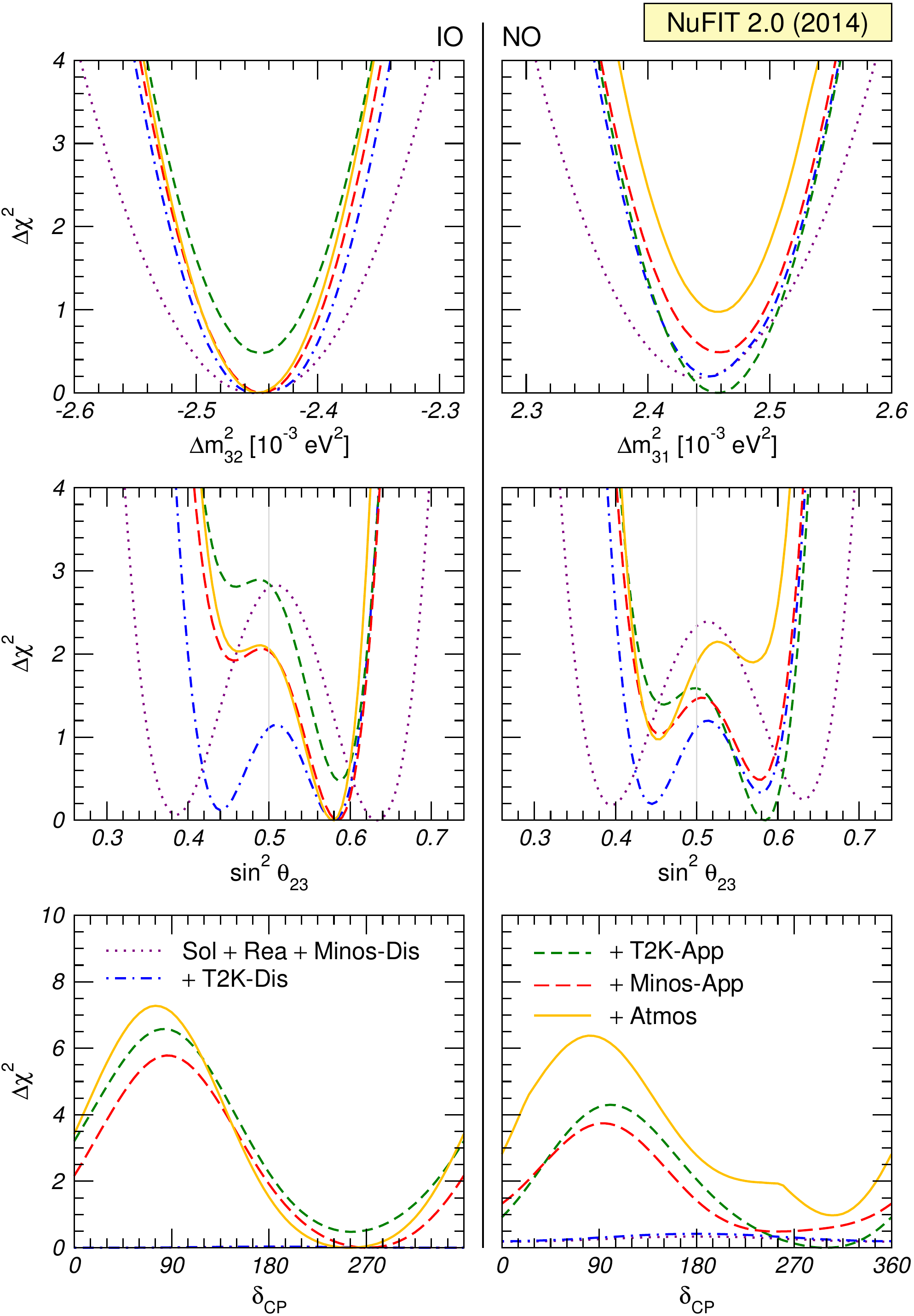}
  \caption{Contribution of different sets of experimental results to
    the present tendencies in the determination of the mass ordering,
    the octant of $\theta_{23}$ and of the CP violating phase. Left
    (right) panels are for IO (NO).  See text for details.}
  \label{fig:chisq-hier}
\end{pagefigure}

All the lines plotted in Fig.~\ref{fig:chisq-hier} include ``by
default'' solar and reactor data, which take care of precisely
determining the undisplayed parameters $\Dmq_{21}$, $\theta_{12}$ and
$\theta_{13}$. To this basic set we progressively add more and more
data, to see how each new piece of information affects the results of
the fit.  Let us then start with the dotted purple (and dot-dashed
blue) curve, which shows the dependence of $\Delta\chi^2$ on the
analysis of solar, reactor and MINOS (plus T2K) $\nu_\mu$ and
$\bar\nu_\mu$ disappearance data.  Being all disappearance experiments
they provide very weak information on $\delta_\text{CP}$, as clearly
visible in the bottom panels. Comparing the minima in the left and
right panels we note a relative difference of $\chi^2(\text{NO}) -
\chi^2(\text{IO}) \sim 0.2$, which means that this combination of data
``favors'' Inverted Ordering by $\sim 0.5\sigma$.  More interestingly,
from the central panels we see that MINOS disappearance data favors a
non-maximal $\theta_{23}$ with $\Delta\chi^2(\theta_{23} = 45^\circ) =
2.8$ ($2.2$) for IO (NO).  Neglecting subleading $\Dmq_{21}$ and
matter effects, the relevant survival probability in MINOS is given by
\begin{equation}
  \label{eq:Pmm}
  P_{\nu_\mu\to\nu_\mu} \approx 1 - \sin^2 2\theta_\text{dis} \,
  \sin^2\frac{\Dmq_{31} L}{4 E_\nu} \,,
  \qquad
  \sin^2\theta_\text{dis} \equiv \sin^2\theta_{23} \cos^2\theta_{13} \,,
\end{equation}
where $L$ is the baseline and $E_\nu$ is the neutrino energy. Hence,
the probability is symmetric under $\theta_\text{dis} \to \pi/2 -
\theta_\text{dis}$.  In the limit $\theta_{13} = 0$ the effective
angle $\theta_\text{dis}$ reduces to $\theta_{23}$, and a preference
for non-maximal $\theta_\text{dis}$ mixing leads to the appearance of
two symmetric minima in the first and second octant of
$\theta_{23}$. Such degeneracy persists also for $\theta_{13} \ne 0$,
and is responsible for the presence of two quasi-degenerate minima at
$\sin^2\theta_{23} = 0.63$ and $0.39$. On the other hand, T2K
disappearance data are better fitted with maximal $\theta_\text{dis}$,
so once they are included in the analysis (dot-dashed blue line) the
positions of the two minima move to values $\sin^2\theta_{23}=0.58$
and $0.44$ while the preference for non-maximal mixing reduces to
$\Delta\chi^2(\theta_{23} = 45^\circ) = 1$.  The comparison of the
dotted purple and dot-dashed blue curves also shows the impact of the
inclusion of T2K disappearance data on the overall determination of
$\theta_{23}$ and $\Dmq_{3\ell}$.

The short-dashed green line shows the effect of further adding to the
analysis the T2K $\nu_e$ appearance data.  First, we see that the
absolute minima now occurs for NO with $\Delta\chi^2(\text{IO}) =
0.6$. In the central panels we see that the quasi-degeneracy of the
octant of $\theta_{23}$ is now broken and the second octant becomes
favored with $\Delta\chi^2(\theta_{23} \leq 45^\circ) = 2.5$ ($1.5$)
for IO (NO).  The lower panels show that after the inclusion of T2K
$\nu_e$ appearance data a minimum appears for
$\delta_\text{CP}=270^\circ$ ($300^\circ$) for IO (NO) with CP
conservation disfavored at $\Delta\chi^2(\sin\delta_\text{CP}=0) =
2.5$ ($1.0$).  This can be understood from the relevant $\nu_e$
appearance probability at T2K and MINOS, which, at the second order in
the small parameters $\sin\theta_{13}$ and $\alpha \equiv \Dmq_{21} /
\Dmq_{31}$ and assuming a constant matter density, takes the
form~\cite{Cervera:2000kp, Freund:2001pn, Akhmedov:2004ny}:
\begin{multline}
  \label{eq:Pme}
  P_{\nu_\mu\to\nu_e}
  \approx 4 \, \sin^2\theta_{13} \, \sin^2\theta_{23}
  \frac{\sin^2 \Delta (1 - A)}{(1 - A)^2} +
  \alpha^2 \sin^2 2\theta_{12} \, \cos^2\theta_{23}
  \frac{\sin^2 A\Delta}{A^2}
  \\
  + 2 \, \alpha \, \sin\theta_{13} \, \sin 2\theta_{12} \,
  \sin2\theta_{23} \cos(\Delta \pm \delta_\text{CP}) \,
  \frac{\sin\Delta A}{A} \, \frac{\sin \Delta (1 - A)}{1 - A} \,,
\end{multline}
with
\begin{equation}
  \Delta \equiv \frac{\Dmq_{31} L}{4 E_\nu} \,,
  \quad A \equiv \frac{2 E_\nu V}{\Dmq_{31}} \,.
\end{equation}
Here $L$ is the baseline, $E_\nu$ is the neutrino energy, and $V$ is
the effective matter potential~\cite{Wolfenstein:1977ue} which for T2K
yields $|A| \sim \text{few}~\%$.  The first term in Eq.~\eqref{eq:Pme}
(which dominates for large $\theta_{13}$) depends on
$\sin^2\theta_{23}$ and therefore is sensitive to the octant.  Reactor
experiments with $L \sim 1$~km, on the other hand, provide a
measurement of $\theta_{13}$ independent of $\theta_{23}$
\begin{equation}
  \label{eq:peereac}
  P_{\nu_e\to\nu_e} = 1 - \sin^2 2\theta_{13}
  \sin^2 \frac{\Dmq_{31} L}{4 E_\nu} + \mathcal{O}(\alpha^2) \,.
\end{equation}
At present the $\nu_e$ appearance results from T2K points towards an
excess with respect to what is expected for the best fit value of
$\sin^2\theta_{13}$ determined by the reactor experiments for maximal
$\theta_{23}$ (\textit{i.e.}, for $2\sin^2\theta_{23} = 1$), hence the
tendency towards the $\theta_{23} > 45^\circ$ minimum.  The matter
effects in Eq.~\eqref{eq:Pme} make this tendency different for NO and
IO, while the last term introduces a $\delta_\text{CP}$ modulation of
the effect.  For fixed $\theta_{13}$ and $\theta_{23}$,
$P_{\nu_\mu\to\nu_e}(\delta_\text{CP}) - P_{\nu_\mu\to\nu_e}(\pi) \geq
0$ ($\leq 0$) for $\delta_\text{CP} \geq \pi$ ($\leq \pi$). For the
best fit values of $\theta_{13}$ and $\theta_{23}$ from the previous
reactor and LBL $\nu_\mu$ disappearance results, the T2K $\nu_e$
appearance signal is better fitted with $\delta_\text{CP}$ values
which enhance the corresponding appearance probability.  Conversely we
see that adding the less significant MINOS $\nu_e$ appearance data in
the analysis (long-dashed red curves) tends to slightly reduce the
size of these effects for NO and it shifts the global minimum from NO
to IO.

Finally the solid orange curves show the impact of including the
atmospheric data in the analysis. Comparing the solid orange and
long-dashed red curves we see that atmospheric data contributes
positively to the significance of the tendency towards IO and
$\delta_\text{CP} > \pi$. While for IO it does not affect the tendency
towards second $\theta_{23}$ octant, for NO it ``shifts'' this
tendency to the first octant. The preference for $\theta_{23} <
45^\circ$ for NO is related to an excess of sub-GeV e-like events, an
effect which has already been discussed since many years (see,
\textit{e.g.},~\cite{Kim:1998bv, Peres:1999yi, Peres:2003wd,
  Gonzalez-Garcia:2004cu}). The fact that this preference is not
visible for IO is probably related to multi-GeV data, which are
affected by matter effects and therefore provides some sensitivity to
the mass ordering. Identifying the relevant bins is difficult, given
the large amount of data points entering the atmospheric fit. We
stress that such effects happen at the level of 1-2 units in $\chi^2$
and hence are not statistically significant.\footnote{In this respect
  it is also important to stress that already since SK2 the
  Super-Kamiokande collaboration has been presenting its experimental
  results in terms of a large number of data samples.  The rates for
  some of those samples cannot be theoretically predicted (and
  therefore included in a statistical analysis) without a detailed
  simulation of the detector, which can only be made by the
  experimental collaboration itself. Hence, although our results
  represent the most up-to-date analysis of the atmospheric neutrino
  data which can be performed outside the collaboration, such an
  analysis has unavoidable limitations. For details on our simulation
  of the data samples and the statistical analysis see the Appendix of
  Ref.~\cite{GonzalezGarcia:2007ib}.}

\begin{figure}\centering
  \includegraphics[width=0.8\textwidth]{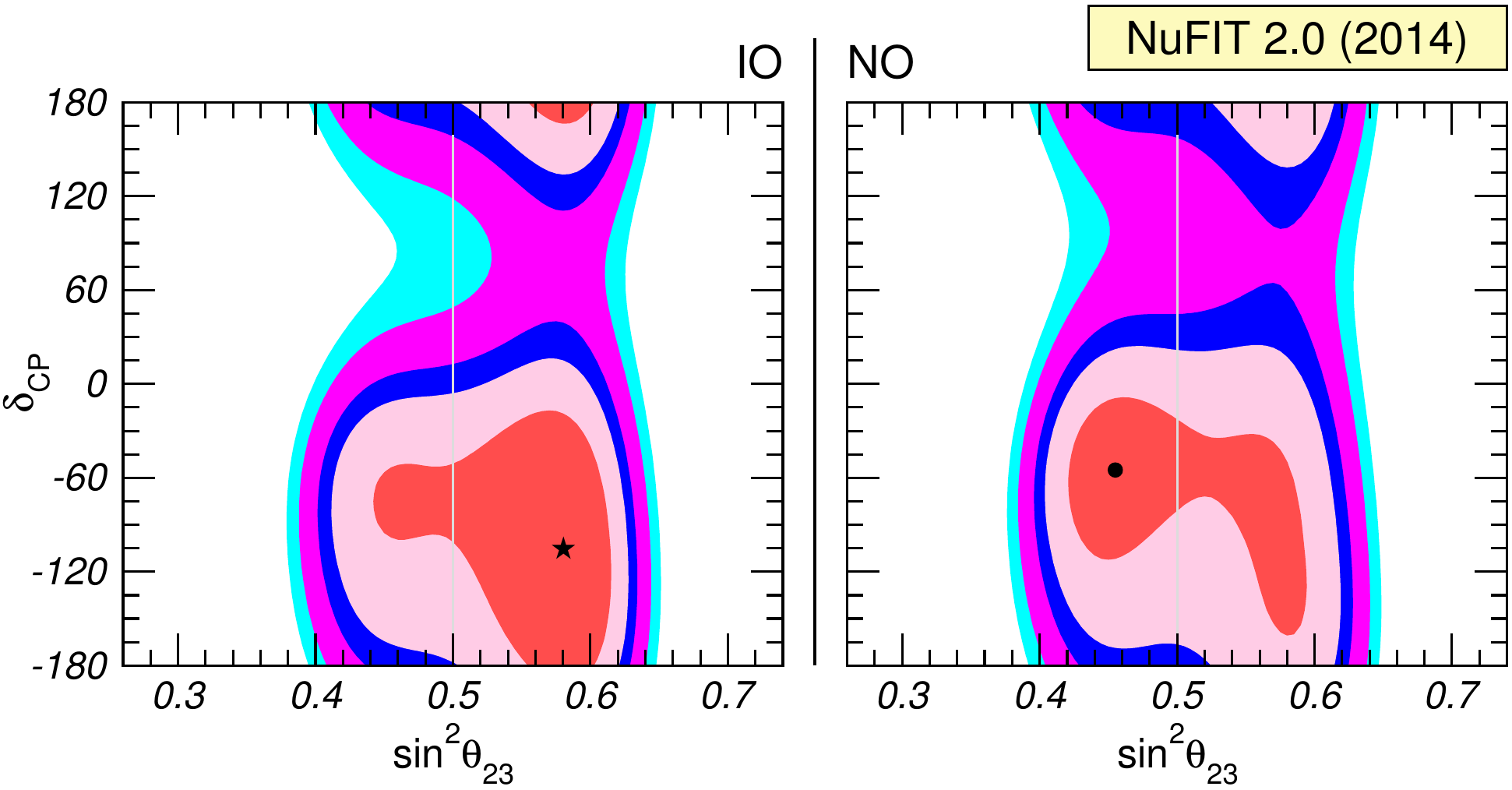}
  \caption{Allowed regions from the global data at $1\sigma$, 90\%,
    $2\sigma$, 99\% and $3\sigma$ CL (2~dof) in the $(\theta_{23},
    \delta_\text{CP})$ plane, after minimizing with respect to all
    undisplayed parameters. The left (right) panel corresponds to IO
    (NO).  Contour regions in both panels are derived with respect to
    the global minimum which occurs for IO and is indicated by a
    star. The local minimum for NO is shown by a black dot.}
  \label{fig:region-hier}
\end{figure}

In order to highlight the pattern of correlations between
$\delta_\text{CP}$ and $\sin^2\theta_{23}$ we show in
Fig.~\ref{fig:region-hier} the allowed regions of the global analysis
projected into the plane of these two parameters. Correlations between
$\delta_\text{CP}$ and other oscillation parameters are mostly trivial
and are therefore omitted.

\subsection{Remarks on confidence levels for $\delta_\text{CP}$}
\label{sec:tenten.stat}

In order to study the information from data on the CP phase we
consider the quantity
\begin{equation}
  \label{eq:Dc}
  \Delta\chi^2(\delta_\text{CP}) \equiv
  \min_{x \neq \delta_\text{CP}} \chi^2(\delta_\text{CP}, x) - \chi^2_\text{min} \,,
\end{equation}
where the first term on the right hand side is minimized with respect
to all oscillation parameters except $\delta_\text{CP}$ ($x =
\theta_{12}, \theta_{13}, \theta_{23}, \Dmq_{21}, \Dmq_{31}$) and the
last term is the $\chi^2$ minimum with respect to all oscillation
parameters. We have shown $\Delta\chi^2(\delta_\text{CP})$ for various
data sets in the lower panels of Fig.~\ref{fig:chisq-hier}, as well as
in the corresponding panel in Fig.~\ref{fig:chisq-glob} for the global
data.  The standard way to derive confidence intervals for
$\delta_\text{CP}$ is to assume that $\Delta\chi^2(\delta_\text{CP})$
follows a $\chi^2$-distribution with 1~dof, and then apply cuts
corresponding to, \textit{e.g.}, $\Delta\chi^2 = 0.99$, $2.71$,
$3.84$, $6.63$ for 68\%, 90\%, 95\%, 99\% CL, respectively. This
procedure relies on Wilks theorem to hold~\cite{Wilks}. However, in
the case of $\delta_\text{CP}$ some of the hypothesis of this theorem
may be violated~\cite{Schwetz:2006md, Blennow:2014sja}.  One reason
for this is the complicated non-linear dependence of the event rates
on $\delta_\text{CP}$. Present sensitivity is so poor, that those
non-linearities (as well as the periodic character of
$\delta_\text{CP}$) become relevant already at very low
CL. Furthermore, parameter degeneracies (especially with
$\theta_{23}$, see discussion in the previous sub-section) affect the
distribution of the test statistic $\Delta\chi^2$ from
Eq.~\eqref{eq:Dc}.

\begin{figure}\centering
  \includegraphics[width=\textwidth]{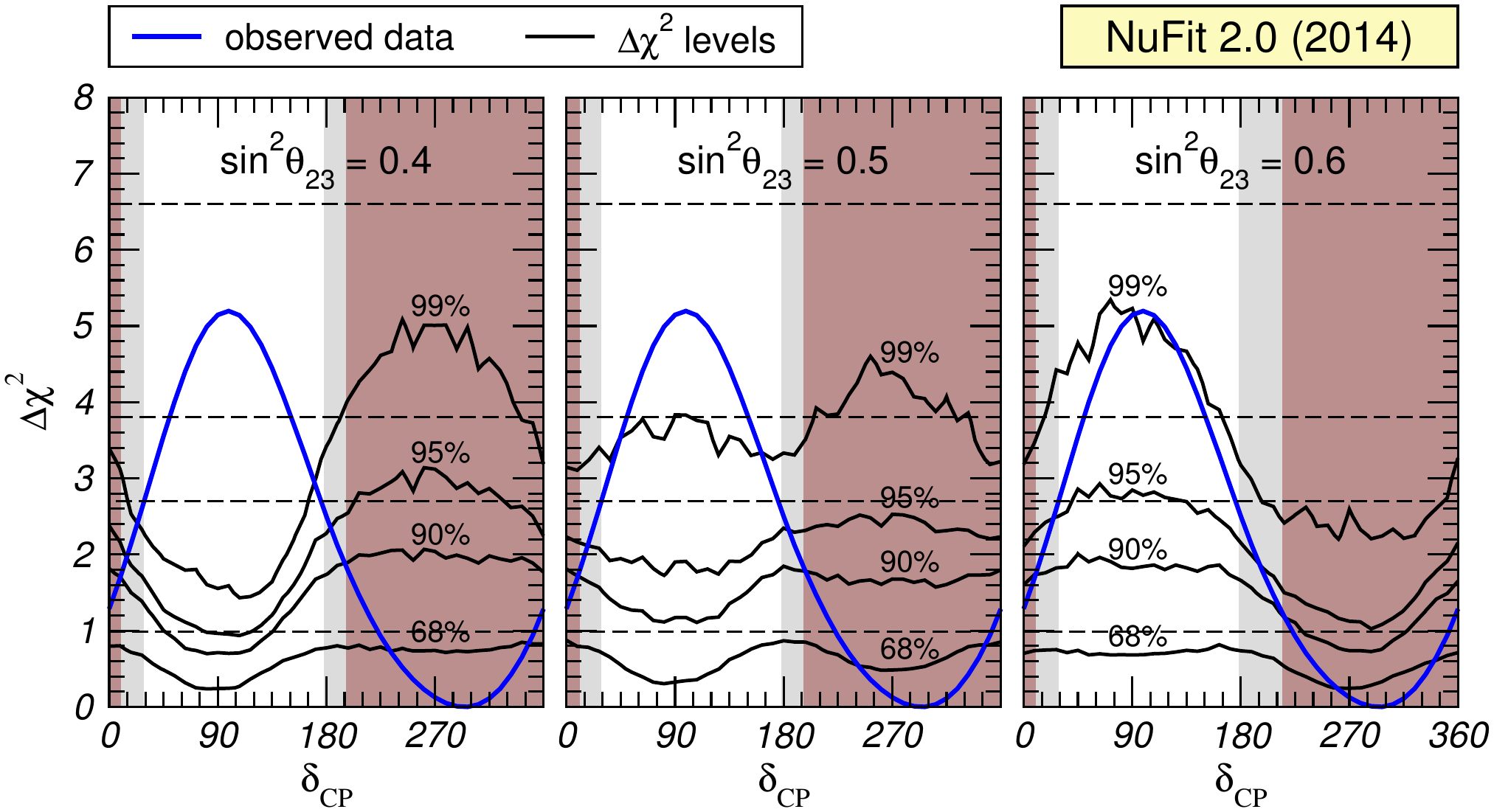}
  \caption{Black curves show the $\Delta\chi^2$ levels corresponding
    to 68\%, 90\%, 95\%, 99\% CL obtained from a Monte Carlo
    simulation of T2K appearance and disappearance data. Dashed lines
    correspond to the canonical values based on the $\chi^2$
    distribution with 1~dof. The blue curve shows the observed
    $\Delta\chi^2$ using T2K data. The shaded regions indicate the
    90\% confidence interval for $\delta_\text{CP}$ based on the
    distribution from simulated pseudo-data (brown) and on the
    $\chi^2$ approximation (gray). The three panels correspond to
    different assumptions on the true value of $\theta_{23}$ used to
    generate the pseudo-data. In the fit all parameters except
    $\delta_\text{CP}$ and $\theta_{23}$ are fixed to the global best
    fit values, assuming normal mass ordering.}
  \label{fig:dchq-levels}
\end{figure}

In order to address such concerns we have performed a Monte Carlo
study of T2K data (appearance and disappearance). We consider a test
statistic similar to the $\Delta\chi^2$ given in Eq.~\eqref{eq:Dc};
however, in order to keep calculation time manageable we fix all
oscillation parameters except $\delta_\text{CP}$ and $\theta_{23}$ to
their best fit values from the global fit assuming normal mass
ordering. Hence, in the notation of Eq.~\eqref{eq:Dc} we have only $x
= \theta_{23}$. In particular, since we keep also $\theta_{13}$ fixed,
the main feature of the complementarity of long-baseline appearance
and medium-baseline reactor data is maintained. We have checked that
allowing $\theta_{13}$ to vary imposing the constraint from Daya Bay
data has a negligible impact on $\Delta\chi^2(\delta_\text{CP})$
compared to fixing it to the best fit value. The resulting
$\Delta\chi^2(\delta_\text{CP})$ is shown as blue curve in
Fig.~\ref{fig:dchq-levels} (identical in all three panels). It differs
somewhat from the global result displayed in
Figs.~\ref{fig:chisq-glob} or~\ref{fig:chisq-hier}, which include more
data, but it captures the essential features and suffices for the
purpose of studying the statistical properties of the test statistic.

In order to estimate the probability distribution for
$\Delta\chi^2(\delta_\text{CP})$ we proceed as follows. We scan the
parameter space of $\delta_\text{CP}$ and $\sin^2\theta_{23}$ and for
a given point of assumed true values we generate a large number of
pseudo-data samples for T2K. For each data set we calculate the value
of the test statistic $\Delta\chi^2(\delta_\text{CP})$ and in this way
we obtain a distribution for it. We scan 41 points in
$\delta_\text{CP}$ and 3 points for $\sin^2\theta_{23}$, and for each
of those points we generate 5000 pseudo-data samples. The black curves
in Fig.~\ref{fig:dchq-levels} show the values of
$\Delta\chi^2(\delta_\text{CP})$ which are larger than 68\%, 90\%,
95\% and 99\% of all generated data samples. We observe quite large
deviations from the corresponding values based on the
$\chi^2$-distribution for 1~dof, shown by the dashed lines in the
figure. Interestingly we find also a rather strong dependence on the
assumed true value of $\theta_{23}$.

The behavior of the curves can be understood qualitatively. Due to the
non-linearity of $\delta_\text{CP}$ (its cyclic nature) and the poor
sensitivity mentioned above it actually counts as less than 1 full
degree of freedom, which implies distributions more concentrated at
lower values than the $\chi^2$-distribution for 1~dof, as observed in
Fig.~\ref{fig:dchq-levels}.  The rather strong variations for
non-maximal values of $\theta_{23}$, including a flipped behavior for
$\delta_\text{CP}$ smaller or larger $\pi$ between $\sin^2\theta_{23}
= 0.4$ and $0.6$ can be understood in terms of a degeneracy.  For
$\theta_{23} < \pi/4$ and $\delta_\text{CP} \sim 3\pi/2$ as well as
for for $\theta_{23} > \pi/4$ and $\delta_\text{CP} \sim \pi/2$ there
is a degeneracy between the two octants of $\theta_{23}$ which
effectively enhances the number of degrees of freedom in the
fit.\footnote{The presence of this degeneracy can be understood from
  Eq.~\eqref{eq:Pme} considered at fixed $\theta_{13}$ and $\Delta
  \simeq \pi/2$ (first oscillation maximum).}

Now we can compare $\Delta\chi^2(\delta_\text{CP})$ obtained from the
observed data to the expected distribution. If the observed
$\Delta\chi^2(\delta_\text{CP})$ is larger than the values obtained
for $x\%$ of the pseudo-data samples for that true value of
$\delta_\text{CP}$ we exclude this value of $\delta_\text{CP}$ at the
$x\%$~CL. In Fig.~\ref{fig:dchq-levels} we show as an example the
resulting 90\% confidence interval for $\delta_\text{CP}$ as brown
shaded area. This corresponds to the confidence interval according to
the Feldman-Cousins (FC) prescription~\cite{Feldman:1997qc}. It has to
be compared to the corresponding interval based on the
$\chi^2$-approximation, indicated by the gray area in the plot.

We can draw the following conclusions from the exercise shown in
Fig.~\ref{fig:dchq-levels}:
\begin{enumerate}
\item the confidence intervals based on the Monte Carlo simulation are
  smaller than the ones based on the $\chi^2$-approximation. Hence,
  the latter is conservative;

\item for confidence levels $\lesssim 90\%$ the confidence intervals
  are similar, whereas for higher confidence levels differences become
  significant. In particular, at 99\%~CL all values of
  $\delta_\text{CP}$ are allowed using the $\chi^2$-approximation,
  whereas a region around $\delta_\text{CP}\sim \pi/2$ remains
  excluded by the 99\%~CL FC interval;

\item the CL with which $\delta_\text{CP}\sim \pi/2$ can be
  disfavored depends strongly on the unknown true value of
  $\theta_{23}$. For $\sin^2\theta_{23} = 0.6$, $\delta_\text{CP}
  \simeq \pi/2$ is excluded at about 99\%~CL, whereas for
  $\sin^2\theta_{23} = 0.4$ it is excluded at very high CL. In all
  cases, the CL based on the Monte Carlo is higher than in the
  $\chi^2$-approximation which again can be considered conservative.
\end{enumerate}

Let us conclude this section by commenting that ideally such a
simulation should be performed also for the global
analysis. Unfortunately this is currently out of question, in
particular due to atmospheric neutrino data, which is very
computational intensive and does play a non-negligible role in the
global fit for $\Delta\chi^2(\delta_\text{CP})$, see
Fig.~\ref{fig:chisq-hier}.  However, we believe that the above results
based on T2K are approximately representative also for the global fit.
One may expect that, with more statistics, distributions become more
close to the expected $\chi^2$-distribution. However, preliminary
estimates indicate that parameter degeneracies may lead to deviations
also in a high-statistics scenario.

%%%%%%%%%%%%%%%%%%%%%%%%%%%%%%%%%%%%%%%%%%%%%%%%%%%%%%%%%%%%%%%%%%%%%%%%%%%%%%

\section{Summary}
\label{sec:summary}

We have presented the results of an updated (as of summer 2014) global
analysis of solar, atmospheric, reactor and accelerator neutrino data
in the framework of three-neutrino oscillations.  Quantitatively the
present determination of the oscillation parameters is listed in
Table~\ref{tab:results}, and the corresponding leptonic mixing matrix
is given in Eq.~\eqref{eq:umatrix}. From the present analysis we have
derived the maximum allowed CP violation in the leptonic sector as
parametrized by the Jarlskog determinant, $J_\text{CP}^\text{max} =
0.0329 \pm 0.0009$ ($\mathrel{\pm} 0.0027$) at $1\sigma$ ($3\sigma$).
All these results have also been shown in terms of unitarity triangles
in Fig.~\ref{fig:region-viola} which further illustrate the ability of
global oscillation data to obtain information on leptonic CP
violation.

The global analysis presents a series of tensions between data sets as
well as some 1-2$\sigma$ effects in the determination of less known
parameters ($\theta_{23}$, mass ordering, and $\delta_\text{CP}$)
which we denote as ``tendencies'' and we discuss in
Sec.~\ref{sec:tenten}. We can summarize these results as follows:
\begin{itemize}
\item due to the very precise determination of the flux-independent
  near-far ratio from Daya Bay and RENO, the so-called reactor
  neutrino anomaly (\textit{i.e.}, the tension between the predicted
  reactor fluxes in Refs.~\cite{Mueller:2011nm, Huber:2011wv} and the
  event rates observed in short-baseline reactor experiments) results
  only in a $0.5\sigma$ uncertainty on the determination of
  $\theta_{13}$;

\item the long-standing $\sim 2\sigma$ tension between the best fit
  values of $\Dmq_{21}$ as determined from the analysis of KamLAND and
  solar data is still unresolved. This tension is driven by both the
  indication of a non-zero day-night effect at SK, and by the lack of
  evidence of a low energy turn-up in the \Nuc{8}{B} energy spectrum
  as measured by SNO, SK4 and Borexino.  In both cases the $\Dmq_{21}$
  value favored by KamLAND is in disagreement with the expectations
  from the standard LMA-MSW solution;

\item the uncertainty on the determination of $\Dmq_{21}$ and
  $\theta_{12}$ due to the choice of Standard Solar Model associated
  with the ``solar composition problem'' is negligible;

\item at present the precision on the determination of
  $|\Dmq_{3\ell}|$ from $\nu_\mu$ disappearance experiments (mainly
  T2K and MINOS) is comparable to that from $\nu_e$ disappearance
  experiments (\textit{i.e.} reactor experiments including, in
  particular, the spectral information from Daya Bay);

\item for Inverted Ordering, the ``tendency'' towards non-maximal
  mixing and second octant of $\theta_{23}$ is driven mainly by two
  effects: (a) the non-maximality favored by MINOS $\nu_\mu$
  disappearance, and (b) the ``mismatch'' between the best fit
  $\theta_{13}$ obtained from $\bar\nu_e$ disappearance at reactors
  and from $\nu_\mu \to \nu_e$ at T2K. Atmospheric results do not
  alter this;

\item for Normal Ordering, such preference for non-maximal
  $\theta_{23}$ mixing is considerably weaker than for IO; also, in
  this case the global best-fit occur in the first $\theta_{23}$
  octant, mostly driven by atmospheric data;

\item the ``mismatch'' between reactor and T2K results is the driving
  effect in the present dependence of the global $\Delta\chi^2$ on the
  CP violating phase with a best fit value close to $\delta_\text{CP}
  = \frac{3}{2}\pi$. Inclusion of the atmospheric results adds
  positively to this effect for both orderings;

\item the tendency towards IO or NO in the present analysis does not
  seem to result from any consistent effect and it shifts in sign
  depending on the data sets considered.
\end{itemize}

Finally in Sec.~\ref{sec:tenten.stat} we have addressed the issue of
the ``gaussianity'' of the confidence levels attributed to
$\Delta\chi^2(\delta_\text{CP})$ by performing a Monte Carlo study of
T2K data, and we have compared the resulting probability distribution
to that of a $\chi^2$-distribution as usually assumed. Deviations are
expected due to the cyclic nature of $\delta_\text{CP}$ and to the
presence of parameter degeneracies.  The conclusion is that, within
the present data, the use of the $\chi^2$-distribution approximation
is slightly conservative in the determination of the excluded range of
$\delta_{\text CP}$ at confidence levels $\gtrsim 90\%$. The
differences however are not very significant as illustrated in
Fig.~\ref{fig:dchq-levels}.

Future updates of this analysis will be provided at the website quoted
in Ref.~\cite{nufit}.

\section*{Acknowledgments}

We would like to thank M.\ Smy for pointing out to us the importance
of the day-night asymmetry for the tension between the solar and
KamLAND fits, and H.\ Nunokawa for spotting a typo in
Eq.~\eqref{eq:jmax}.  T.S.\ also thanks Mattias Blennow, Pilar Coloma
and Enrique Fernandez-Martinez for extensive discussions on the
results presented in section~\ref{sec:tenten.stat}.  This work is
supported by Spanish MINECO grants FPA2012-31880, FPA2012-34694 and
FPA2013-46570, by the Severo Ochoa program SEV-2012-0249 and
consolider-ingenio 2010 grant CSD-2008-0037, by CUR Generalitat de
Catalunya grant 2009SGR502, by USA-NSF grant PHY-09-69739 and
PHY-13-16617, and by EU grant FP7 ITN INVISIBLES (Marie Curie Actions
PITN-GA-2011-289442).

\appendix

\section{List of data used in the analysis}
\label{sec:appendix}

\subsection*{Solar experiments}

\begin{itemize}
\item Chlorine total rate~\cite{Cleveland:1998nv}, 1 data point.

\item Gallex \& GNO total rates~\cite{Kaether:2010ag}, 2 data points.

\item SAGE total rate~\cite{Abdurashitov:2009tn}, 1 data point.

\item SK1 full energy and zenith spectrum~\cite{Hosaka:2005um}, 44
  data points.

\item SK2 full energy and day/night spectrum~\cite{Cravens:2008aa}, 33
  data points.

\item SK3 full energy and day/night spectrum~\cite{Abe:2010hy}, 42
  data points.

\item SK4 1669-day energy spectrum and day/night
  asymmetry~\cite{sksol:nu2014}, 24 data points.

\item SNO combined analysis~\cite{Aharmim:2011vm}, 7 data points.

\item Borexino 740.7-day low-energy data~\cite{Bellini:2011rx}, 33
  data points.

\item Borexino 246-day high-energy data~\cite{Bellini:2008mr}, 6 data
  points.
\end{itemize}

\subsection*{Atmospheric experiments}

\begin{itemize}
\item SK1--4 (including SK4 1775-day) combined
  data~\cite{skatm:nu2014}, 70 data points.
\end{itemize}

\subsection*{Reactor experiments}

\begin{itemize}
\item KamLAND combined DS1 \& DS2 spectrum~\cite{Gando:2010aa}, 17
  data points.

\item CHOOZ energy spectrum~\cite{Apollonio:1999ae}, 14 data points.

\item Palo Verde total rate~\cite{Piepke:2002ju}, 1 data point.

\item Double Chooz 227.9-day spectrum~\cite{Abe:2012tg}, 18 data
  points.

\item Daya Bay 621-day spectrum~\cite{db:nu2014}, 36 data
  points.

\item RENO 800-day near \& far total rates~\cite{reno:nu2014}, 2
  data points (with free normalization).

\item SBL reactor data (including Daya-Bay total flux at near
  detector), 77 data points~\cite{Declais:1994ma,
Kuvshinnikov:1990ry, Declais:1994su, Vidyakin:1987ue, Vidyakin:1994ut,
Kwon:1981ua, Zacek:1986cu, Greenwood:1996pb, Afonin:1988gx, db:nu2014}.
\end{itemize}

\subsection*{Accelerator experiments}

\begin{itemize}
\item MINOS $10.71\times 10^{20}$~pot $\nu_\mu$-disappearance
  data~\cite{Adamson:2013whj}, 39 data points.

\item MINOS $3.36\times 10^{20}$~pot $\bar\nu_\mu$-disappearance
  data~\cite{Adamson:2013whj}, 14 data points.

\item MINOS $10.6\times 10^{20}$~pot $\nu_e$-appearance
  data~\cite{Adamson:2013ue}, 5 data points.

\item MINOS $3.3\times 10^{20}$~pot $\bar\nu_e$-appearance
  data~\cite{Adamson:2013ue}, 5 data points.

\item T2K $6.57\times 10^{20}$ pot $\nu_\mu$-disappearance
  data~\cite{Abe:2014ugx}, 16 data points.

\item T2K $6.57\times 10^{20}$ pot $\nu_e$-appearance
  data~\cite{Abe:2013hdq}, 5 data points.
\end{itemize}

\bibliographystyle{JHEP}
\bibliography{references}

\end{document}